\documentstyle[12pt,epsf]{article}

\def\beq{\begin{equation}}
\def\eeq{\end{equation}}
\def\eq{\end{equation}}
\def\ba{\begin{eqnarray}}
\def\ea{\end{eqnarray}}

\def\centeron#1#2{{\setbox0=\hbox{#1}\setbox1=\hbox{#2}\ifdim
\wd1>\wd0\kern.5\wd1\kern-.5\wd0\fi
\copy0\kern-.5\wd0\kern-.5\wd1\copy1\ifdim\wd0>\wd1
\kern.5\wd0\kern-.5\wd1\fi}}
\def\ltap{\;\centeron{\raise.35ex\hbox{$<$}}{\lower.65ex\hbox{$\sim$}}\;}
\def\gtap{\;\centeron{\raise.35ex\hbox{$>$}}{\lower.65ex\hbox{$\sim$}}\;}
\def\gsim{\mathrel{\gtap}}
\def\lsim{\mathrel{\ltap}}

\def\slepton{\tilde{\ell}}
\def\sneutrino{\tilde{\nu}}
\def\selectron{\tilde{e}}
\def\smuon{\tilde{\mu}}
\def\stau{\tilde{\tau}}
\def\higgsino{\widetilde{H}}
\def\wino{\widetilde{W}}
\def\bino{\widetilde{B}}

\def\msusy{\widetilde{m}}
\def\amususy{a_{\mu}^{\rm SUSY}}
\def\amuexp{a_{\mu}^{\rm exp}}
\def\amusm{a_{\mu}^{\rm SM}}

\def\amususytau{a_{\mu}^{{\rm SUSY}-\tilde{\tau}}}
\def\amsusytau{a_{\mu}^{{\rm SUSY}-\tilde{\tau}}}
\def\amususymu{a_{\mu}^{{\rm SUSY}-\tilde{\mu}}}
\def\amu{a_{\mu}}

\def\desusystau{d_e^{{\rm SUSY}-\tilde{\tau}}}
\def\desusyse{d_e^{{\rm SUSY}-\tilde{e}}}

\def\pslash{\!{\not \! p}}
\def\pslashb{\!{\not \! \bar{p}}}

\def\tanb{\tan \beta}

\def\mwino{m_{\widetilde{W}}}
\def\mbino{m_{\widetilde{B}}}
\def\mmu{m_{\mu}}
\def\mtau{m_{\tau}}
\def\amususy{a_{\mu}^{\rm SUSY}}
\def\amu{a_{\mu}}
\def\amuexp{a_{\mu}^{\rm exp}}
\def\amusm{a_{\mu}^{\rm SM}}

\def\dijLL{(\delta^{\ell}_{ij})_{LL}}
\def\dijRR{(\delta^{\ell}_{ij})_{RR}}

\def\dijRL{(\delta^{\ell}_{ij})_{RL}}

\def\dabLL{(\delta^{\ell}_{12})_{LL}}
\def\dacLL{(\delta^{\ell}_{13})_{LL}}
\def\dbcLL{(\delta^{\ell}_{23})_{LL}}
\def\dabRR{(\delta^{\ell}_{12})_{RR}}

\def\dcaLL{(\delta^{\ell}_{31})_{LL}}
\def\dcbLL{(\delta^{\ell}_{32})_{LL}}
\def\dbaRR{(\delta^{\ell}_{21})_{RR}}
\def\dcaRR{(\delta^{\ell}_{31})_{RR}}
\def\dcbRR{(\delta^{\ell}_{32})_{RR}}

\def\eLij{\epsilon_{Lij}}
\def\eRij{\epsilon_{Rij}}
\def\eLba{\epsilon_{L21}}
\def\eRba{\epsilon_{R21}}
\def\eLca{\epsilon_{L31}}
\def\eRca{\epsilon_{R31}}
\def\eLcb{\epsilon_{L32}}
\def\eRcb{\epsilon_{R32}}

\def\phac{\varphi_{1331}}

\newcommand{\newc}{\newcommand}
\newc{\qbar}{{\overline q}}
\newc{\Kahler}{K\"ahler }
\newc{\deltaGS}{\delta_{\rm GS}}
\begin{document}

\begin{titlepage}

\begin{center}
\vspace*{-1cm}
\hfill   SCIPP-01-09 \\
\hfill SU-ITP-01-05 \\
\hfill hep-ph/0104254 \\
\vskip .6in
{\Large \bf Supersymmetric Relations Among} \\
\vspace{.1in}
{\Large \bf Electromagnetic Dipole Operators}
\vskip 0.4in
{\bf Michael Graesser$^{a}$}~ and ~{\bf Scott Thomas$^{b}$ }

\vskip 0.15in

$^a${\em Department of Physics\\
University of California, Santa Cruz, CA 95064}

\vspace{.1in}

$^b${\em Department of Physics\\
Stanford University, Stanford, CA 94305}
\end{center}
\vskip 0.15in

\baselineskip=14pt

\begin{abstract}

Supersymmetric contributions to all leptonic electromagnetic
dipole operators have essentially identical diagramatic structure.
With approximate slepton universality this allows the muon
anomalous magnetic moment to be related to the electron electric
dipole moment in terms of supersymmetric phases, and to
radiative flavor changing lepton decays in terms of small violations of
slepton universality.  If the current discrepancy between the
measured and Standard Model values of the muon anomalous
magnetic moment is due to supersymmetry, the current bound
on the electron electric dipole moment then implies that the phase
of the electric dipole operator is less than $2 \times 10^{-3}$.
Likewise the current bound on $\mu \to e \gamma$ decay
implies that the fractional selectron--smuon mixing in the left--left
mass squared matrix, $\delta m_{\smuon \selectron}^2 / m_{\slepton}^2$,
is less than $10^{-4}$.  These relations and constraints are
fairly insensitive to details of the superpartner spectrum for
moderate to large $\tan \beta$.

\end{abstract}
\end{titlepage}

\baselineskip=16pt

\newpage

High precision measurements of low energy processes can often
provide useful probes of physics beyond the Standard Model.
Many processes of this type involve the coupling of a photon to
Standard Model fermions, such as anomalous magnetic dipole
moments \cite{oldmuon},
electric dipole moments (EDMs) \cite{oldedm}, and rare radiative
decays \cite{oldflavor}.
The effective operators which describe these interactions
are all of the electromagnetic dipole form.
The magnitude of these operators in general depends on the
overall scale and details of the heavy particle
mass spectrum as well as the interactions which violate
the requisite symmetries.

In supersymmetric theories the one-loop contributions to all
the electromagnetic dipole
operators have very similar diagramatic structure.
In this %letter
paper we point out that in the lepton sector
this similarity allows
%the overall magnitude of supersymmetric contributions
%to the dipole operators
%to be related to the muon anomalous magnetic moment.
%This allows
the muon anomalous
magnetic moment to be related to the electron EDM in terms
of the phase of the electromagnetic dipole operator,
and to the rate for radiative lepton decays,
$\mu \rightarrow e \gamma$,
$\tau \rightarrow \mu \gamma$ and
$\tau \rightarrow e \gamma$,
in terms of violations of slepton universality.
For moderate to large $\tan \beta$ and with approximate
slepton universality these relations turn
out to be fairly insensitive to details of the superpartner
mass spectrum.
If the discrepancy between the current measured value of
the muon anomalous magnetic moment \cite{bnl} and the Standard Model
value is interpreted as arising from supersymmetry,
fairly model independent bounds can be obtained on the
phase of the dipole operator and fractional flavor violating
splitting in the slepton mass squared matrix from
the current experimental bounds on the electron EDM
and $\ell_i \rightarrow \ell_j \gamma$ decays respectively.
Alternately, an upper limit on the supersymmetric contribution to
the muon anomalous magnetic moment provides a lower
limit for the most stringent possible bounds arising from
the electron EDM and  $\ell_i \rightarrow \ell_j \gamma$
decays.
The relations among the supersymmetric electromagnetic
dipole operators presented here are particularly interesting
and useful in light of the recent high precision measurement of the
muon anomalous magnetic moment \cite{bnl}.
The relation between the muon anomalous magnetic moment and 
radiative flavor changing lepton decays has previously been 
considered in the context of supersymmetric see-saw models
of neutrino masses \cite{seesaw}.

In the next section %\ref{opsection}
the structure of the various supersymmetric
contributions to electromagnetic dipole operators
are illustrated.
The relative importance of various classes of diagrams
are presented and the dominant contribution identified.
Under the assumption of slepton universality and
proportionality the lepton dipole operators for different
generations are shown to be related by ratios of the lepton
masses.
The manner in which these ratios may be modified by slepton
sflavor violation is also described.
In section \ref{muonsection}
supersymmetric contributions to the muon anomalous
magnetic moment are discussed, and the magnitude of possible
contributions from sflavor violation evaluated.
These contributions are shown generally to be smaller than
the flavor conserving contributions.
% enough
%so as not to spoil the relation of the muon anomalous
%magnetic moment to the other electromagnetic dipole operators.
The relationship between the
electron EDM and muon anomalous magnetic moment
is discussed in section \ref{edmsection}.
For moderate to large $\tan \beta$ it is shown that the electron EDM
is dominated by a single supersymmetric phase to lowest order
in gaugino--Higgsino mixing and assuming strict gaugino unification.
Under the assumption of slepton universality and proportionality,
the most stringent
possible model independent limit on this phase
arising from the $^{205}$Tl EDM experiment consistent with the
Brookhaven muon $g-2$ experiment is derived.
Contributions to the electron EDM from sflavor violation are
also considered, and in particular,
important contributions from staus
are identified.
Radiative flavor changing lepton decays arising from
transition dipole moments are considered in section
\ref{decaysection}.
Fairly model--independent stringent
bounds
on sflavor violating mass
squared
mixings implied by current limits on $\ell_i \rightarrow \ell_j \gamma$
and consistent with the Brookhaven muon $g-2$ experiment
are derived. The manner in which they are model--independent
are described below.
The complete expressions for the one-loop
supersymmetric contributions to electromagnetic
dipole operators in the large $\tan \beta$ limit and
to first order in gaugino--Higgsino mixing
are given in the appendix.

%The two important issues are: more data and
%better control of the theory prediction. Marciano.
%Two different techniques for extracting the hadronic
%contribution to the photon self--energy are not in
%complete concordance. Still, it is
%due to Jegerlener the theory and experimental measurements
%remain roughly 2 sigma apart.

%Other non--supersymmetric sources may also
%explain the discrepancy.
%For large extra dimensions the theory is not under control,
%and at best one can obtain an estimate of the minimal
%correction one might expect \cite{mg}. Other sources
%are...refer to recent papers, Marciano.

% ---------------------------------------------------

\section{Electromagnetic Dipole Operators}
\label{opsection}

The coupling of an on-shell Dirac fermion to the electromagnetic
field strength may be represented by the Lagrangian dipole
operator
\beq
- {1 \over 2} {\cal D}_f~ \overline{f}_L \sigma^{\mu \nu}
 f_R F_{\mu \nu}
 - {1 \over 2} {\cal D}_f^* ~\overline{f}_R \sigma^{\mu \nu}
 f_L F_{\mu \nu}
\label{dipoleop}
\eeq
where ${\cal D}_f$ is the dipole moment coefficient,
and $f_{L,R} = P_{L,R} f$ are the left- and right-handed
chiral components of the Dirac fermion.
The dipole moment coefficient can in general be complex and
have non-trivial flavor structure.
The on-shell dipole operator is chirality violating and so must
vanish with the fermion mass.
Supersymmetric contributions are therefore proportional
to the fermion mass, and suppressed by two powers
of the superpartner mass scale, rendering
the dipole operator effectively dimension six.

Supersymmetric contributions to anomalous magnetic moments
require only violation of chiral symmetry, which is
already violated in the Standard Model by the fermion Yukawa couplings.
In contrast, electric dipole moments require in addition
the violation of parity and time-reversal symmetries beyond that
of the Standard Model.
And flavor changing radiative decays require violation of
flavor symmetries, which in the leptonic sector are not
violated within the Standard Model.
The strategy here is therefore to use the similarity of
the one-loop supersymmetric contributions discussed below
to determine the overall scale of the dipole operators
from anomalous magnetic moments and to relate this to electric dipole moments
in terms of supersymmetric phases which violate parity and
time reversal, and to radiative flavor changing decays in terms
of supersymmetric violations of flavor symmetries.
Experimentally, only the muon anomalous magnetic moment is measured
sufficiently accurately to allow the possibility of
discerning the supersymmetric contribution.
As a practical matter, model independent relations among the dipole operators
are therefore only useful in the leptonic sector.
Under the mild assumptions of approximate universality and proportionality
discussed below, these relations turn out to be rather insensitive
to details of the superpartner mass spectrum.

In order to determine the dominant diagrams and
relations among the dipole operators it is instructive
to consider the parametric dependences of various contributions.
The supersymmetric one--loop
contributions to the electromagnetic
dipole operators arise from virtual sleptons and charginos or
neutralinos.
These can be classified according to whether
the one-particle-irreducible diagrams are chirality
conserving or violating.
The chirality conserving diagrams give dimensions six
operators which reduce to the chirality
violating dipole operator (\ref{dipoleop})
on-shell through the external equations of motion.
The chirality violating diagrams give the effective
dimension six dipole operator (\ref{dipoleop}) directly.

The chargino and neutralino mass eigenstates are general
mixtures of the Bino, Wino, and Higgsino interaction
eigenstates.
However, for $m_i^2 - \mu^2 \gg m_Z^2$, which holds over much
of the parameter space, the mixing may be treated perturbatively
in $m_Z^2 / (m_i^2 - \mu^2)$, where $m_i =\{ m_{\bino},m_{\wino} \}$
are the Bino and Wino Majorana mass parameters and
$\mu$ is the Higgsino Dirac mass parameter.
It is therefore sufficient to work to lowest non-trivial
order in gaugino-Higgsino mixing.
It is seen below
that diagrams with gaugino-Higgsino mixing
can provide the most
important contributions to the dipole moments.
In addition,
for lepton dipole operators it is sufficient to
work to first order in the small lepton Yukawa coupling.
This greatly simplifies classification of the supersymmetric
diagrams.

\begin{figure}[ht]
\begin{center}
\epsfxsize= 2.19 in
\leavevmode
%\epsfbox[98 505 518 597]{dipolec.ps}
\epsfbox[217 504 401 596]{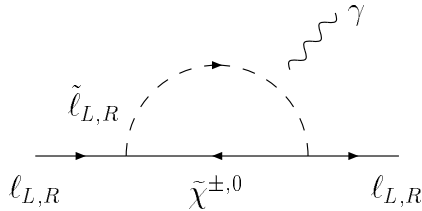}
\end{center}
%\vspace{.2in}
\caption[f1]{Chirality conserving left--left or right--right
contributions
to lepton electromagnetic dipole operators.
Arrows indicate the flow of fermion or scalar-partner chirality.
%Internal fermions are neutralinos or charginos.
%A cross indicates a chirality violating propagator.
%Internal scalars are sleptons.  A dot indicates
%a left-right slepton propagator or equivalently a
%left-right mass squared mixing insertion.
To lowest order in the fermion Yukawa coupling
only gaugino--gaugino propagators contribute to the
internal chargino or neutralino propagators.
The external photon is attached to internal charged
lines in all possible ways.}
\label{figc}
\end{figure}

The chirality conserving one--loop diagrams for leptons
%arising from virtual sleptons and charginos or neutralinos
are shown in Fig. \ref{figc}.
Since the use of the external lepton equation of motion
to obtain the dipole operator (\ref{dipoleop}) involves
the lepton Yukawa coupling it is sufficient consider
diagrams which do not include additional powers of the
lepton Yukawa.
Left--right slepton mixing must vanish with the lepton mass,
and so can be ignored in the chirality conserving diagrams.
Likewise, the charginos and neutralinos only couple through
gaugino components to lowest order since the Higgsino components
couple through the lepton Yukawa.
The chargino propagator is therefore given by
$\langle \wino^+ \wino^- \rangle$ where throughout
a sum over mass eigenstates is understood.
Charginos only contribute to the left--left diagram.
For the neutralino diagrams
Bino--Wino mixing arises only at second order in gaugino--Higgsino
mixing and so the chirality conserving
neutralino propagators are given predominantly by
$\langle \wino^{0*} \wino^0 \rangle$ and
$\langle \bino^* \bino \rangle$ which contribute to the
left--left and right--right diagrams respectively.
The parametric dependence of the chirality conserving
contributions arising from the diagrams of Fig. \ref{figc}
are
\beq
\chi~{\rm Conserving}~~RR~~\langle \bino^* \bino \rangle ~~~:~~~~
{\cal D}_f \sim {g_1^2~m_{\ell} \over 16 \pi^2 ~\msusy^2}
\label{cRR}
\eq
\beq
\chi~{\rm Conserving}~~LL~~\langle \wino^{+,0*} \wino^{-,0}
    \rangle ~~~:~~~~
{\cal D}_f \sim {g_2^2~m_{\ell} \over 16 \pi^2 ~\msusy^2}
\label{cLL}
\eq
where $16 \pi^2$ represents the loop factor,
$\msusy$ represents the superpartner mass scale
determined by the heaviest particle in the loop,
and the lepton mass $m_{\ell}$ arises from lepton equation
of motion.
The Bino contributions are suppressed by
a factor $g_1^2 / g_2^2=\tan^2 \theta_w$
compared with Wino.
The dominant chirality conserving contribution then arises
for the left--left diagram through Wino propagators.

\begin{figure}[ht]
\begin{center}
\epsfxsize= 5 in
\leavevmode
%\epsfbox[98 505 518 597]{dipolev.ps}
\epsfbox[98 505 518 597]{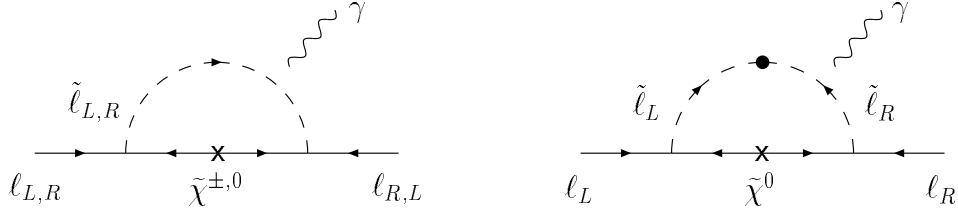}
\end{center}
%\vspace{.2in}
\caption[f1]{Chirality violating left--right contributions
to lepton electromagnetic dipole operators.
Arrows indicate the flow of fermion or scalar-partner
chirality.
%Internal fermions are neutralinos or charginos.
A cross indicates a chirality violating propagator.
%Internal scalars are sleptons.
A dot indicates a left--right slepton propagator or equivalently a
left--right mass squared mixing chiral insertion.
To lowest order in the fermion Yukawa coupling
only gaugino--Higgsino propagators contribute
to the diagram without a chiral insertion on the scalar line,
and only gaugino--gaugino propagators to the diagram
with a chiral insertion on the scalar line.
The later are dominated by Bino--Bino propagators up to
second order in gaugino-Higgsino mixing.
The external photon is attached to internal charged
lines in all possible ways.}
\label{figv}
\end{figure}

The chirality violating one--loop diagrams for leptons
%arising from virtual sleptons and charginos or neutralinos
are shown in Fig. \ref{figv}.
These diagrams contribute directly to the operator (\ref{dipoleop})
and have an explicit factor of the lepton Yukawa coupling.
In the first class of diagrams the lepton Yukawa
coupling, $\lambda_{\ell}$,
arises directly in the vertex which couples the down-type
Higgsino components
of the chargino or neutralino
to the slepton and external lepton.
To lowest order in the lepton Yukawa the other vertex then
couples to the gaugino component of the chargino or
neutralino proportional to a gauge coupling.
Left--right slepton mixing can also be neglected in these
diagrams to lowest order in the lepton Yukawa.
The chargino propagator in these diagrams is then
$\langle \wino^+ \higgsino^-_d \rangle$.
Since the Wino couples to left-handed fields, only the
left handed slepton arises in this class of chargino diagrams.
The neutralino propagators are
$\langle \wino^0 \higgsino^0_d \rangle$ which arises only with
the left-handed slepton and
$\langle \bino \higgsino^0_d \rangle$ which arises only with
the right-handed slepton.
All these propagators require gaugino--Higgsino mixing
which arises through coupling with the Higgs condensate
proportional to a gauge coupling.
To lowest order in mixing the
$\langle \wino^+ \higgsino^-_d \rangle$
and $\langle \wino^0 \higgsino^0_d \rangle$
chirality violating propagators
are proportional to $m_{\wino}(g_2 v_u)\mu$
through mixing of chirality violating Wino and Higgsino
propagators through the up-type Higgs condensate,
and $g_2 v_d$
through mixing of chirality conserving propagators through the
down-type Higgs condensate.
The chirality violating
$\langle \bino^0 \higgsino^0_d \rangle$ propagator
is likewise proportional to $m_{\bino}(g_1 v_u)\mu$ and $g_1 v_d$.
%For moderate to large $\tan \beta = v_u/v_d$ all these
%propagators
%diagrams are enhanced by the gaugino--Higgsino mixing
Mixings through the down-type Higgs condensates are
suppressed for moderate to large $\tan \beta = v_u/v_d$.
The parametric dependence of this class of chirality
violating contributions coming from the first diagram
of Fig. \ref{figv} in this case are
\beq
\chi~{\rm Violating}~~LR~~\langle \bino \higgsino^0_d \rangle ~~~:~~~~
{\cal D}_f \sim {g_1^2~m_{\bino} \mu~ m_{\ell} \tan \beta \over
    16 \pi^2 ~\msusy^4}
\label{vBH}
\eq
\beq
\chi~{\rm Violating }~~LR~~\langle \wino^{+,0}
       \higgsino^{-,0}_d \rangle ~~~:~~~~
{\cal D}_f \sim {g_2^2~m_{\wino} \mu~ m_{\ell} \tan\beta
    \over 16 \pi^2 ~\msusy^4}
\label{vWH}
\eq
where $m_{\ell} \sim \lambda_{\ell} v_d$.
Even though these contributions
require gaugino--Higgsino mixing, they
are parametrically enhanced by
a factor of $\tan \beta$ with respect to the chirality
conserving contributions (\ref{cRR}) and (\ref{cLL})
because of the coupling to the up-type Higgs condensate.

The second class of chirality violating diagrams involve
the lepton Yukawa coupling through left--right scalar
mixing.
To first order in the lepton Yukawa this mixing may
be treated as a mass squared insertion on the slepton
propagator.
Under the assumption of proportionality of the scalar
tri-linear soft $A$-terms,
the left--right mixing mass squared for sleptons
is given by
$(A - \mu \tan \beta) m_{\ell}$.
The factor of $\mu \tan \beta$ arises from a superpotential
cross term between the Higgsino mass parameter and lepton
Yukawa coupling which mixes the scalar sleptons
through the up-type Higgs condensate.
Left--right mixing only occurs for charged sleptons.
So only
neutralinos contribute to the diagrams with
external charged leptons.
To lowest order in the lepton Yukawa only the gaugino
components of the neutralinos couple the sleptons to the
external leptons proportional to a gauge coupling.
Only the Bino component of the neutralinos couples
to the right-handed slepton.
And since $\langle \wino^0 \bino \rangle$ arises only at
second order in gaugino--Higgsino mixing the chirality violating
neutralino propagators are given predominantly
by $\langle \bino \bino \rangle$.
For moderate to large $\tan \beta$ the parametric dependence
of this class of chirality violating contributions
coming from the second diagram of Fig. \ref{figv}
is
\beq
\chi~{\rm Violating}~~LR~~\langle \bino \bino \rangle ~~~:~~~~
{\cal D}_f \sim {g_1^2~m_{\bino}\mu ~m_{\ell} \tan \beta
   \over 16 \pi^2 ~\msusy^4}
\label{vBB}
\eq
where $m_{\bino}$ arises from the chirality violating
Bino propagator and $\mu \tan \beta$ from left--right
slepton mixing.
These contributions are also enhanced by
a factor of $\tan \beta$ with respect to the chirality
conserving contributions (\ref{cRR}) and (\ref{cLL}).
% again
%because of the coupling to the up-type Higgs condensate.

For moderate to large $\tan \beta$ the chirality violating
contributions to dipole operators should dominate
over the chirality conserving ones because of the coupling
to the up-type Higgs condensate.
Among these, the neutralino
contributions (\ref{vBH})
and (\ref{vBB}) which involve Bino coupling(s)
are suppressed
by a factor $(g_1^2/g_2^2)m_{\bino} / m_{\wino}$
compared with the chargino and neutralino
contributions (\ref{vWH}) which involve Wino couplings.
With gaugino unification $m_{\bino}/m_{\wino} \simeq g_1^2/g_2^2$
which implies an overall suppression of
$(g_1^4 / g_2^4) = \tan^4 \theta_w$.
So with slepton universality and proportionality
the first diagram of Fig. \ref{figv},
which includes a left-handed slepton and Wino coupling,
should give the dominant contributions to the dipole operator
for moderate to large $\tan \beta$ \cite{moroi,carena}.
This expectation is useful in identifying the microscopic
phase bounded by the electron EDM discussed in section
\ref{edmsection}, and the source of slepton flavor violation
bounded by $\ell_i \rightarrow \ell_j \gamma$ decays
discussed in section \ref{decaysection}.

Even though the neutralino and chargino diagrams of
the dominant chirality violating contribution are parametrically
identical, the loop integrals differ because the external
photon does not couple to the same internal lines.
For equal superpartner masses the neutralino diagram
turns out to be a factor 6 smaller than the chargino \cite{moroi}.
For general superpartner masses the relative importance of these two
diagrams depends on the ratios
$x \equiv \mu ^2 / m^2_{\tilde{W}}$ and
$y \equiv m^2_{\tilde{e}_L} / m^2_{\tilde{W}}$.
An explicit evaluation of these
diagrams (see the appendix for details) indicates that
the relative importance of the chargino
to neutralino diagram strictly
increases for large left-handed slepton masses,
$i.e.~,$ for $y>1,x=1$, or for small $\mu$,
$i.e.~,$ for $y=1,x<1$, and strictly decreases in
the other directions in parameter space, that is
for small left-handed slepton masses,
$i.e. ~,$ for $y<1,x=1$, or large
$\mu$, $i.e. ~,$ for $y=1,x>1$. Even though the ratio is decreasing
for $y<1,x=1$, the chargino diagram is more important for
well--motivated values for $y$.
For example,  with $y=1/10$ the chargino diagram is three times larger
than the neutralino diagram,
and for the extreme value of $y=1/100$ it is still twice as large.
There is a similar behavior for
$y=1$ with $x$ variable.
As noted above, for $x<1$ the chargino diagram is
even more important than for the case of equal superpartner masses.
For $x>1$ the ratio decreases
rather slowly, with the chargino diagram still four times as
large even at the extreme value of $x=100$.
We therefore conclude that the chargino first diagram of Fig. \ref{figv}
is dominant over the neutralino first diagram
over essentially all of parameter space.

%**************
%A%nd the sum of all the chirality violating
%neutralino diagrams of Fig. \ref{figv} is a factor
%$-(1-\tan^2 \theta)/6 \simeq -0.12$
%smaller than the chirality violating chargino diagram of
%Fig. \ref{figv} in this case \cite{moroi}.
%The dominance of the chargino diagram for moderate to large
%$\tan \beta$ with slepton proportionality extends over most of parameter
%space \cite{carena,moroi} in the absence of slepton flavor violation.
%**************

The importance of the
second neutralino
diagram of Fig. 2 depends on the size
of the $\mu$ parameter and the right--handed
slepton mass compared to the
other superpartner masses. %\cite{moroi}.
For equal superpartner masses is a factor of
$3 g_2^2 / g_1^2 =3/ \tan^2 \theta \simeq 10$
smaller than the dominant chargino diagram discussed above.
However, in the limit in which the
$\mu$ term is much larger than all the other superpartner
masses this diagram is easily seen to be larger
than the other gaugino--Higgsino diagrams. %\cite{moroi}.
The reason
is that the neutralino diagram with the left--right mass
insertion is directly proportional to $\mu$, whereas
the diagrams with gaugino--Higgsino propagators decouple
at least as fast as $\mu ^{-1}$. This behavior is
preserved even if the left-handed slepton and
Wino masses are large and comparable to $\mu$,
whereas the right-handed slepton and Bino masses remain small.
In this case the
chargino diagram decouples as
\beq
{\cal D}^{\chi^+} _{\mu} \rightarrow -\frac{e g^2_2}{64 \pi^2}
\frac{m_{\mu}^2}{m^2 _{\tilde{L}}} \tan \beta  ,
\eeq
whereas in the same limit the neutralino second diagram
of Fig. \ref{figv} with
the left--right mass squared insertion decouples more slowly,
\beq
{\cal D}^{\chi^0_{LR}} _{\mu} \rightarrow -\frac{e
g^2_1}{96 \pi^2}
\frac{m^2_{\mu} \mu m_{\tilde{B}}}{m^2_{\tilde{L}}
m^2_{\tilde{R}} } \tan \beta
\label{nlr}
\eeq
and for $\mu \simeq m_{\tilde{e}_L}$ this
can easily dominate. As
$\mu$ is decreased further the chargino diagram
becomes dominant.
For $\mu$ small compared to the Wino  and left-handed
slepton mass
the chargino diagram decouples as $m^{-3}_{\tilde{L}}$, but
the loop integral has an logarithmic infra-red divergence
that is cutoff by $\mu$. In this limit the chargino
diagram is
\beq
{\cal D}^{\chi^+} _{\mu} \rightarrow -\frac{e g^2_2}{16 \pi^2}
\frac{m_{\mu}^2 \mu}{m^3 _{\tilde{L}}} \left(\ln \left(
\frac{m^2_{\tilde{L}}}{\mu ^2} \right) -\frac{11}{6} \right)
\tan \beta
\eeq
whereas the neutralino diagram with the left--right mass
insertion is unchanged from (\ref{nlr}). In this limit
the
logarithm is large, and
the chargino diagram still dominates
this neutralino diagram
for left--handed slepton and wino masses
more than roughly 10 times larger
than the $\mu$ term, Bino, and right--handed slepton masses.

%For $\mu$ term and
%left-handed slepton and wino masses large compared to
%the bino and right-handed sletpn mass the neutralino
%diagram wins. Otherwise the chargino diagram is more important.

Finally, the importance of the first neutralino diagram of
Fig. 2 with Bino coupling and right-handed sleptons,
rather than Wino coupling and left-handed sleptons considered
above, obviously
depends on $\mu$ and
the right-handed and Bino mass spectrum. If gaugino
unification is assumed,
$m_{\bino} / m_{\wino} = g_1^2 / g_2^2$,
then this neutralino diagram
and the chargino diagram depend on the mass ratios
$x \equiv \mu^2 /m^2_{\tilde{W}}$,
$y\equiv
m^2_{\tilde{e}_L} /m^2_{\tilde{W}}$, and
$z \equiv m^2_{\tilde{e}_R} /m^2_{\tilde{W}}$. The
relative importance of these two diagrams is
determined by these ratios.
%We set
Consider the case with
$m_{\tilde{e}_L} = m_{\tilde{W}}$, $i.e.$, $y=1$.,
% and vary the other two ratios.
For comparable
right-handed and left-handed slepton masses
the chargino diagram is dominant. For example,
for $z=1$ corresponding to $m_{\tilde{e}_R}=m_{\tilde{W}}$,
the bino neutralino diagram is typically a factor of roughly 10
smaller than the dominant chargino diagram for $\mu$ in the range
$1/9 < x <9$.
For a right-handed slepton
mass small compared to the Wino mass the neutralino
diagram may however be comparable to the chargino diagram.
More concretely, for $z=1/9$ corresponding to $m_{\tilde{e}_R}
\simeq m_{\tilde{B}}=m_{\tilde{e}_L}/3$,
the neutralino diagram is typically less than but
comparable to the chargino diagram for various values of
$\mu$ or $x$; the
neutralino to chargino
diagram ratio for $x=1/9$ is 0.7, and steadily
decreases to 0.5 for $x=9$.

For even lighter right-handed slepton masses this neutralino
diagram remains larger than the chargino diagram by a factor
of a few.
It might be imagined that in this limit
the neutralino diagram scales
as an inverse power of the right-handed slepton mass
and then easily dominates.
In fact, there is no infra-red divergence.
This is apparent in the effective theory below the
Bino and Higgsino masses.
At this scale integrating out the Bino and Higgsino generates
a dimension five operator which couples two
fermions and two sleptons.
This operator contributes at one-loop to the muon dipole
operator, and by inspection
naively vanishes as the right-handed slepton
mass goes to zero. The loop integral however is linearly
divergent and this cancels the vanishing mass dependence to
leave a finite part.
In this limit the Bino neutralino first diagram of Fig. \ref{figv}
approaches
\beq
{\cal D}^{\chi ^0 _R} _{\mu} \rightarrow
\frac{e g^2_1}{32 \pi^2} \frac{m_{\mu}}
{\mu m_{\tilde{B}}} .
\eeq
in this limit.
Thus the heavier Bino and Higgsino masses set the scale for this
contribution rather than the
lighter right-handed slepton mass.
Comparing this result to the chargino diagram assuming
gaugino unification and equal $\mu$ term Higgsino mass,
Wino mass, and left-handed slepton mass shows that
in these limits it is larger
by a factor of two than the chargino diagram.

The result is that within the gaugino mass unification
assumption the first Bino neutralino
diagram of Fig. 2 with right-handed sleptons
is actually more important than the
chargino
diagram for a right-handed slepton mass roughly three times smaller
than the mass scale of the $SU(2)_L$ superpartner
masses. For heavier
right-handed slepton masses the chargino diagram is dominant.
Finally, it is interesting to note that
for equal superpartner masses and assuming slepton proportionality
there is an accidental cancelation between the leading
$\tan \beta$ contributions to the
Bino neutralino first diagram of Fig. \ref{figv} and the second
Bino neutralino diagram of Fig. \ref{figv} which involves a
left--right mass squared insertion.

Independent of which diagrams dominate,
up to very small corrections proportional to powers
of the lepton Yukawa coupling,
all contributions to the lepton dipole operators
are proportional to a single power of the
lepton mass, as discussed above.
With slepton universality and proportionality of the scalar
tri-linear soft terms this implies that the dipole operators
for different leptons are related simply by ratios
of the Yukawa couplings or equivalently lepton masses.
This applies diagram by diagram.
For example, for the electron and muon
\beq
{\cal D}_{e} \simeq { m_e \over m_{\mu}} {\cal D}_{\mu}
\label{relation}
\eeq
for both the real and imaginary parts, and likewise
for the tau dipole operator.
This relation is good over all of parameter space with
slepton universality and proportionality.
This relation will be used in subsequent sections to relate
the muon anomalous magnetic moment to the electron EDM
in terms of the phase of the operator, and to radiative
$\ell_i \rightarrow \ell_j \gamma$ decays in terms
of small violations of slepton flavor.

Violations of slepton universality and proportionality
can in principle modify the relation (\ref{relation}).
%in interesting ways.
The magnitude of the scalar tri-linear
$A$-terms for the first two generations
are in principle limited only by the requirement
that the radiatively induced contribution to the
lepton mass not be larger than the observed
lepton masses \cite{soft}.
For non-proportional $A \gsim \mu \tan \beta$ the relation
(\ref{relation}) would be modified.
However, in almost all theories of supersymmetry breaking
in which the lepton masses arise in a conventional fashion
from tree-level superpotential
Yukawa couplings, the $A$-terms are at most of order
of the other supersymmetry breaking mass parameters.
Splittings of the $\selectron$, $\smuon$, and $\stau$
masses would also of course modify the precise
relation (\ref{relation}).
Such splittings depend on the underlying theory of flavor
and supersymmetry breaking and in most models are
small at least for the first two generations.

More interesting modifications of the relation (\ref{relation})
can arise from sflavor violation in the slepton
soft mass squared matrix.
For the first two generations,
sflavor violating mixings can introduce dependence on
a heavier lepton mass.
This occurs in the second chirality violating
diagrams of Fig. \ref{figv}.
%which include left--right slepton mixing proportional
%to a lepton mass.
%For example, with selectron--stau mixing the
%electron dipole operator
Sflavor violation in the slepton propagators
%connecting the interaction
%vertices to the left--right mass squared insertion
allows left--right mass squared insertions proportional
to $m_{\mu}$ or $m_{\tau}$ for the electron dipole operator,
and $m_{\tau}$ for the muon.
For moderate to large $\tan \beta$ the parametric dependence
of this class of chirality violating contributions to the
electron and muon dipole operators is
\beq
{\cal D}_e \sim {g_1^2 ~ m_{\bino} \mu \over
     16 \pi^2 \msusy^4}
\left[ \dabLL \dbaRR~\mmu +
        \dacLL \dcaRR~\mtau  \right] \tanb
\label{desflavor}
\eq
\beq
{\cal D}_{\mu} \sim {g_1^2 ~ m_{\bino} \mu \over
     16 \pi^2 \msusy^4}
\left[ \dbcLL \dcbRR~\mtau \right] \tanb ~~~~~~~~~~~~~~~~~~~~~~~~~~
\label{Dmuflavor}
\eq
where throughout
\beq
(\delta^{\ell}_{ij})_{LL} \equiv
   {\delta m^2_{\slepton_{iL} \slepton_{jL}}
   \over m^2_{\slepton_L} }
\label{deltadef}
\eeq
represents insertions of sflavor violating left--left mass squared
mixings in the slepton propagators,
and likewise for right--right and left--right sflavor
violating mass squared terms.
%Sflavor violation also is of course important ..
The potential importance of these sflavor violating mixing effects
in introducing dependence on heavier fermion masses
depends on the magnitude of the sflavor violation and
on the specific dipole operator.
Possible contributions associated to the EDM and
flavor changing operators are presented in
subsequent sections.

\section{Muon Anomalous Magnetic Moment}
\label{muonsection}

The anomalous magnetic moment
$a_f \equiv (g-2)/2$ of a Dirac fermion is related to the
dipole operator (\ref{dipoleop}) by
\beq
a_f = {2 |m_f| \over e Q_f}
   |{\cal D}_f| \cos \varphi %\left( {\rm Arg}({\cal D}_f) \right)
\label{afdef}
\eeq
where $Q_f$ the fermion electric charge,\footnote{The electron
electric charge is $Q_e=-1$.} and in a general basis
\beq
\varphi \equiv  {\rm Arg}({\cal D}_f m_f^*)
\label{relphase}
\eq
is the relative phase between the
dipole operator and fermion mass.

Supersymmetric contributions to the anomalous magnetic
dipole moments of various fermions can be related
in terms of the microscopic parameters of the theory.
As discussed in section \ref{opsection},
%since all the supersymmetric contributions are
slepton universality and proportionality imply that
the lepton dipole moment operators are related by
ratios of lepton
masses (\ref{relation}).
With the definition (\ref{afdef}) this gives the
well known relation relation that supersymmetric
contributions to anomalous magnetic moments are related
by ratios of fermion masses squared.
For example, for the electron and muon
\beq
a_{e}^{\rm SUSY} \simeq ~{m_e^2 \over \mmu^2} ~ \amususy
\label{arelation}
\eq
and likewise for the tau.

The best measured anomalous magnetic moment
in proportion to the fermion mass squared is for the muon.
%It is for this reason that sflavor conserving contributions
%to the muon magnetic moment are the best probes of the
%supersymmetric mass scale.
%As is discussed below however, sflavor violating contributions
%may enhance the dipole moments by a factor of the tau mass,
%which is proportionally larger for the electron dipole moment.
%But even in this case the electron magnetic moment does not
%provide a useful bound \cite{marciano}. The reason is
%that currently the electron magentic moment provides the
%best measurement of the fine structure constant,
%and any discrepancy between measurement and
%theory may be used to bound new physics only if
%a different measurement for the fine structure
%constant is used, such as the
%quantum hall effect which has a larger error.
Bounds on, or measurements of, $a_{\mu}$ therefore provide
the most useful information about the overall magnitude
of supersymmetric contributions to lepton dipole operators.
The Brookhaven muon $g-2$ experiment
has observed a value which differs by at the
2.6 $\sigma$ level from the Standard Model
prediction
$
a_{\mu}^{\rm exp} - a_{\mu}^{\rm SM} = 43 \pm 16\times 10^{-10}
%\label{dis}
$ \cite{bnl}
Additional data and a run utilizing anti-muons will reduce both
statistical and systematic errors.
The largest theoretical uncertainty in the Standard Model prediction
arises from hadronic
contributions to photon vacuum polarization.
At present there is not a
complete concordance among the various theoretical calculations
used to extract the photon polarization from $e^+e^- \rightarrow$ hadrons
and tau decays.
It is of course very important that this uncertainty be better
understood \cite{marciano,badhadron}.

One possible explanation for the experimental
discrepancy \cite{bnl} %(\ref{dis})
is the
existence of additional non-Standard Model contributions to the
muon anomalous magnetic moment.
Electroweak scale supersymmetry can easily give contributions to $\amu$
of the requisite magnitude \cite{recent}.
The dominant chargino--sneutrino diagram
gives a contribution to the muon anomalous magnetic moment of
\beq
a_{\mu}^{\chi^{\pm}} \simeq \frac{
g^2_2 \tan \beta}{32 \pi^2}
\frac{m^2_{\mu}}
{\msusy^2} ~{\rm sgn}(\mu)
\eeq
where $\msusy$ is the effective mass of the virtual superpartners.
This supersymmetric contribution can account for the discrepancy
with $\msusy \sim 50 \sqrt{\tan \beta} ~\hbox{GeV}$
and ${\rm sgn}(\mu) =+$.
The total supersymmetric contribution of course depends on details
of the superpartner mass spectrum and couplings and is model dependent.
However, as detailed in the next two sections,
the relation among supersymmetric contributions
to electromagnetic dipole operators
in terms of violations of time reversal and parity or sflavor symmetries
is not particularly sensitive to details of the superpartner spectrum
and is fairly model independent.
The discrepancy, %(\ref{dis}),
$\amuexp- \amusm$,
if interpreted as arising from supersymmetry,
may therefore be used to set the overall
scale for supersymmetric contributions to
all lepton electromagnetic dipole operators.
Alternately, the discrepancy may be interpreted as an upper limit
on the overall scale of supersymmetric contributions to lepton
dipole operators.

The muon anomalous magnetic moment, or equivalently the dipole moment
coefficient, may be used
to set the scale for other lepton dipole operators
through the relations (\ref{relation}) and
(\ref{arelation}).
Since these relations assume slepton universality and proportionality
it is important to consider the magnitude of possible modifications
of these relations from violations of universality or proportionality.
As mentioned at the end of section {\ref{opsection}, dependence
of a dipole operator on a heavier lepton mass can
be introduced by slepton flavor violation.
For the muon dipole operator the presence of both
right--right and left--left smuon--stau mixing
gives a contribution (\ref{Dmuflavor}) proportional to $\mtau$
through flavor conserving left--right stau mixing in the 
chirality violating Bino second
diagram of Fig. \ref{figv}.
In terms of insertions this corresponds to smuon--stau
mixing insertions on both the left and right handed slepton
lines and a left--right stau mixing insertion proportional to $m_{\tau}$
represented by a dot in the second diagram of Fig. \ref{figv}.
For moderate to large $\tan \beta$
the parametric dependence
of the sflavor violating contribution (\ref{Dmuflavor})
arising from stau mixing
proportional to $m_{\tau}$ in the Bino diagram,
compared to the %$\amususymu$ from the chargino--smuon
dominant flavor conserving contribution (\ref{vWH}) is
%in the limit of universality and proportionality, is
\beq
{ \amsusytau \over \amususymu } \simeq
 {g_1^2 \over  3 g_2^2} {\mtau \over m_{\mu} }
  {m_{\bino} \over m_{\wino} }~
  \dbcLL \dcbRR ~ {h_0 \over f_{+}} h^{\prime \prime}_{0,LR}
\label{naiveest}
\eeq
where
the sflavor violating mixing masses squared
are treated as insertions, and here it is understood
that the flavor violating insertions refer to the real
parts only.
%\beq
%f_{+}=f_{+}(m_{\widetilde{\chi}^{+}_i}^2,m_{\slepton_L}^2)    ~,
%\eq
%\beq
%h_0=h_0(m_{\widetilde{\chi}^0_i}^2,m_{\slepton_L}^2,m_{\slepton_R}^2)
%\eq
The functions $f_+$ and $h_0$
are loop functions defined in the appendix
for the chargino first diagram of
Fig. \ref{figv} and the neutralino second diagram
of Fig. \ref{figv}, and normalized to unity for equal superpartner
masses.
The dimensionless derivative function
\beq
h^{\prime \prime}_{0,LR}
\equiv { m^2_{\slepton_L} m^2_{\slepton_R} \over h_0}
{\partial^2 ~h_0 \over \partial m^2_{\slepton_L}
                    \partial m^2_{\slepton_R} }  =
{1\over h_0}
{ \partial^2 h_0 \over \partial \ln m^2_{\slepton_L}
                    \partial \ln m^2_{\slepton_R}}
\label{dlogfcn}
\eeq
represents the modification of the loop function induced by
the two sflavor violating mixing insertions such that
$h_0 h^{\prime \prime}_{0,LR}$ is the loop function for the stau
contribution $\amususytau$.
This function does not
differ significantly from unity and is fairly insensitive
to details of the superpartner mass spectrum
since it is a logarithmic derivative of the loop function.
The ratio of loop functions $h_0/f_{+}$ does however
depend on the superpartner spectrum.
For equal sparticle masses $h_0/f_{+}=1$ %\cite{moroi},
and $h^{\prime \prime}_{0,LR}=2/5$.
%******************
With gaugino unification $m_{\bino} / m_{\wino} = g_1^2 / g_2^2$,
the ratio (\ref{naiveest}) is then roughly ${\cal O}(10^{-1}-1)
\times (\delta ^{\ell}_{23})_{LL} (\delta
^{\ell}_{23})_{RR}$.
So a significant sflavor violating supersymmetric contribution to the 
muon anomalous magnetic moment would require essentially maximal 
smuon--stau mixing in both the left--left and right--right 
channels. 

The magnitude of possible sflavor violating mixings is 
bounded by radiative flavor changing lepton decays,
as discussed in section \ref{decaysection}.
The bounds depend on the overall magnitude of the electromagnetic
dipole operators, which as discussed here may be related 
to the muon anomalous magnetic moment. 
First, assume that $\amususy$ is dominated by the flavor 
conserving chargino contribution, and that 
the discrepancy $\amuexp - \amususy$ \cite{bnl} 
is interpreted as arising 
from supersymmetry. 
In this case limits on $\tau \rightarrow \mu \gamma$ 
radiative decay discussed 
in section \ref{decaysection} imply that 
$\dbcLL \dcbRR \lsim 10^{-1}$ up to model dependent 
ratios of loop functions. 
So possible sflavor violating contributions to $\amususy$ 
are subdominant in this case. 
If supersymmetric contributions to the muon anomalous magnetic
moment are in fact smaller than the current discrepancy 
$\amuexp - \amususy$ then the bound on sflavor violating 
mixings are weakened since the overall magnitude
of all electromagnetic dipole operators is smaller. 
To estimate the importance of this effect consider 
the Brookhaven muon $g-2$ experiment which may reach an ultimate 
sensitivity of 
$\Delta a_{\mu}^{\rm exp} \sim 4 \times 10^{-10}$ \cite{bnl2}.
If agreement with an improved calculation of the Standard 
Model contribution were obtained at this level then the 
bound on $\amususy$ would improve by approximately an 
order of magnitude. 
This would weaken the bounds obtained in section \ref{decaysection}
derived under the assumption that the 
sflavor conserving chargino contribution dominates $\amususy$
to roughly $\dbcLL \dcbRR \lsim 1$ again up to model dependent
ratios of loop functions. 
In this case the sflavor violating stau--Bino contribution could 
be at most comparable to the 
flavor conserving smuon--chargino contribution. 

So we conclude that for any value of $\amususy$ which could 
be accessible to the ultimate sensitivity of the Brookhaven 
muon $g-2$ experiment, sflavor violating stau contributions
to $\amususy$ are at most comparable to the flavor conserving 
contribution (which would require that both left--left and right--right 
smuon--stau mixing are near maximal), 
and in fact are an order of magnitude smaller 
if the current discrepancy $\amuexp - \amusm$ 
\cite{bnl} is due to supersymmetry. 
This allows $\amususy$ to be identified with the dominant
sflavor conserving chargino contribution for moderate to large 
$\tan \beta$ over most of parameter space. 
And in turn the overall scale for other electromagnetic dipole operators 
may then be related to $\amususy$ through the relations 
(\ref{relation}) and (\ref{afdef}).

\section{Electron Electric Dipole Moment}
\label{edmsection}

An electric dipole moment (EDM) coupling the spin of
a fermion to the electric field is
odd under both parity and time-reversal.
The dipole operator (\ref{dipoleop}) in general violates both
these symmetries and
%In a basis in which the fermion mass is real,
is related to the electric dipole moment by
\beq
d_f = |{\cal D}_f| \sin \varphi % \left( {\rm Arg}({\cal D}_f \right)
\eeq
where $\varphi$ is the relative phase (\ref{relphase})
between the dipole operator coefficient and fermion mass.
An EDM requires that this relative phase be non-vanishing.

%The supersymmetric diagrams which
%contribute to a fermion EDM

Under the assumption of slepton universality and proportionality
the electron and muon dipole operator coefficients,
including the phase,
are related by the ratio of masses (\ref{relation}).
The supersymmetric contribution to the electron EDM
may then be related to the supersymmetric contributions
to the muon anomalous magnetic moment by
\ba
d_e^{\rm SUSY}  & \simeq &  - e ~{m_e \over 2 m_{\mu}^2 }~
  a_{\mu}^{\rm SUSY}~\tan \varphi  \nonumber \\
  & \simeq &
  -4.6 \times 10^{-16} ~a_{\mu}^{\rm SUSY}~ \tan \varphi
  ~~e~{\rm cm}
\label{edmrelation}
\ea
This relation is independent of which diagrams dominate the
dipole operator, or any details of the superpartner spectrum.
It is valid over all of parameter space if slepton universality
and proportionality holds.

If the current discrepancy
between $\amuexp$ and $\amusm$ \cite{bnl} is
interpreted as arising from supersymmetry,
$\amususy \sim 42 \times 10^{-10}$,
the current bound on the electron EDM of
$|d_e| < 4 \times 10^{-27}~e~{\rm cm}$ obtained from
$^{205}$Tl \cite{edmbound} along with the relation
(\ref{edmrelation}) can be used to obtain a bound on the
phase of supersymmetric contribution to the dipole operator of
$$
|\tan \varphi| \lsim 2 \times 10^{-3}
\label{phasebound}
$$
Alternately if $\amuexp - \amusm$ is
taken as an upper limit on $\amususy$,
the above bound %on the phase of the dipole operator
can be interpreted as the most stringent possible bound
the $^{205}$Tl EDM experiment places
on the phase of the dipole operator
consistent with the bound on $\amususy$.

The relation of the phase of the dipole operator to the
underlying phases of the supersymmetric Lagrangian
depends in principle on the relative importance of the individual
diagrams. %which contribute to the dipole operator.
As discussed in section \ref{opsection}, for moderate
to large $\tan \beta$ the chirality violating
diagrams of Fig. \ref{figv} are all parametrically larger
by a factor of $\tan \beta$ than the chirality conserving
diagrams of Fig. \ref{figc}.
In order to relate
the phase of the dipole operator in this limit
to underlying supersymmetric phases
it is instructive to determine the origin of the
phase of each chirality violating diagram.
All the supersymmetric phases arise from relevant
terms in the supersymmetric and supersymmetry breaking
Lagrangians, and appear in the
neutralino, chargino, and slepton mass matrices and therefore
propagators after electroweak
symmetry breaking.

Consider first the dominant chargino diagram of Fig. \ref{figv}.
This diagram involves a sneutrino propagator,
which with slepton universality
does not involve a phase.
The chirality violating chargino propagator
may be obtained to lowest order in Wino-Higgsino
mixing by treating the mixing induced by the Higgs condensate
as an insertion.
In Weyl notation this propagator is
\beq
\langle \wino^{+} \higgsino_d^{-} \rangle \simeq
% { i~m_{\wino}^* \over p^2 - |m_2|^2} ~
% { (-i) g_2 v_u^* \over \sqrt{2} } ~
% { i~\mu^* \over p^2 - |\mu|^2}
{ i \pslash(-i g_2 v_d / \sqrt{2} ) i \pslashb ~+~
  i m_{\wino}^* ( -i g_2 v_u^* / \sqrt{2}) i \mu^* \over
 (p^2 - |m_{\wino}|^2)~(p^2-|\mu|^2) }
\label{WHprop}
\eq
where $v_{u,d} \equiv \sqrt{2} \langle H_{u,d}^0 \rangle$ are the up-
and down-type Higgs boson expectation values.
The first term in (\ref{WHprop}) arises from chirality
conserving Wino and Higgsino propagators
connected through the mixing insertion to the down-type
Higgs condensate, while the second arises from the
chirality violating propagators connected through the
up-type Higgs condensate.
%Since the sneutrino propagator does not introduce a phase,
%the phase of the chargino--sneutrino diagram is just the phase of the
%propagator (\ref{WHprop}).
Including the lepton Yukawa,
$\lambda_{\ell}$,
from the Higgsino--lepton--slepton coupling, the dipole operator phase
arising from the first term in (\ref{WHprop}) proportional to the
down-type Higgs condensate is
${\rm Arg}({\cal D}_{\ell})={\rm Arg}(\lambda_{\ell} v_d)$,
while that from the second term
proportional to the up-type Higgs condensate is
${\rm Arg}({\cal D}_{\ell})={\rm Arg}(\lambda_{\ell} m_{\wino}^* \mu^* v_u^*)$.
The physical phase relevant for the EDM is
the relative phase (\ref{relphase}) between the dipole operator and
lepton mass
$\varphi \equiv {\rm Arg}({\cal D}_{\ell} m_{\ell}^*)$.
The phase of the lepton mass in a general basis is
determined by the down-type Higgs boson expectation value
${\rm Arg}(m_{\ell}) = {\rm Arg}(\lambda_{\ell} v_d)$.
%, where $\lambda_{\ell}$ is the lepton Yukawa coupling.
The physical phase arising from the first term in
the propagator (\ref{WHprop})
%along with the lepton Yukawa $\lambda_{\ell}$,
%from the Higgsino--lepton--slepton coupling,
therefore vanishes ${\rm Arg}({\cal D}_{\ell} m_{\ell}^*)=
  {\rm Arg}(\lambda_{\ell} v_d \lambda_{\ell}^* v_d^*)=0$,
and so contributes
only to the magnetic dipole moment.
The magnitude of this contribution is however suppressed
with respect to the second term in the propagator
for large $\tan \beta$, as discussed in section \ref{opsection}.
The physical phase arising from the second term
in the propagator (\ref{WHprop}) along with the lepton Yukawa,
$\lambda_{\ell}$,
from the Higgsino--lepton--slepton coupling
is ${\rm Arg}({\cal D}_{\ell} m_{\ell}^*) =
{\rm Arg}(\lambda_{\ell}^* m_{\wino}^* \mu^* v_u^* \lambda_{\ell}^* v_d^*)
=- {\rm Arg}(m_{\wino} \mu v_u v_d)$.
In the ground state with broken electroweak symmetry
the relative phase of the up- and down-type Higgs condensates
is anti-aligned
with the Higgs up--Higgs down soft mass parameter,
${\rm Arg}(v_u v_d) = - {\rm Arg}(m_{ud}^2)$, where
$V \supset m_{ud}^2 H_u H_d + h.c.$ \cite{phase}.
The phase of the dominant chirality violating chargino--sneutrino
contribution to the lepton EDM to lowest order in
Wino-Higgsino mixing and to leading order in
$(\tan \beta)^{-1}$ is therefore given by the basis independent
combination of phases \cite{phase}
\beq
\varphi \simeq -{\rm Arg}\left(m_{\wino} \mu (m_{ud}^*)^2 \right)
\label{EDMphase}
\eq
%Assuming slepton universality and

Next consider the neutralino first diagram of Fig. \ref{figv}.
With slepton universality the slepton propagator does not involve a phase.
The chirality violating neutralino propagator
diagram receives contributions at lowest order
from both Wino-Higgsino and Bino-Higgsino mixing.
The $\langle \wino^{0} \higgsino_d^{0} \rangle$ propagator is identical
to the propagator (\ref{WHprop}) including phases.
The physical phase of the leading contribution in $(\tan \beta)^{-1}$
is therefore identical to chargino diagram phase (\ref{EDMphase}).
The Bino--Higgsino propagator to lowest order in mixing in Weyl
notation is very similar
\beq
\langle \bino \higgsino_d^{0} \rangle \simeq
% { i~m_{\wino}^* \over p^2 - |m_2|^2} ~
% { (-i) g_1 v_u^* \over \sqrt{2} } ~
% { i~\mu^* \over p^2 - |\mu|^2}
{ i \pslash(-i g_1 v_d / \sqrt{2} ) i \pslashb ~+~
  i m_{\bino}^* ( -i g_1 v_u^* / \sqrt{2}) i \mu^* \over
 (p^2 - |m_{\bino}|^2)~(p^2-|\mu|^2) }
\label{BHprop}
\eq
Applying the same discussion of the relative phases as given above
for the chargino diagram then implies that the basis independent
physical combination
of phases arising at leading order in $(\tan \beta)^{-1}$
from this diagram is
\beq
\varphi \simeq -{\rm Arg}\left(m_{\bino} \mu (m_{ud}^*)^2 \right)
%\label{EDMphase}
\eq
With strict gaugino unification ${\rm Arg}(m_{\bino}) = {\rm Arg}(m_{\wino})$.
So in this case the phase of this contribution is also identical
that of the chargino diagram (\ref{EDMphase}).

Finally, consider the neutralino second diagram of Fig. \ref{figv}.
The slepton propagator includes left--right mixing
which in a general basis can involve a phase.
To lowest order the left--right mass squared mixing this may
be treated as an insertion in the slepton propagator.
With slepton universality and proportionality
\beq
\langle \slepton_L \slepton_R^* \rangle \simeq
{  i~ \left[ i \lambda_{\ell} ( A v_d - \mu^* v_u^*) \right] ~i
  \over  (p^2 - m_{\slepton_L}^2) (p^2 - m_{\slepton_R}^2) }
\label{sleptonprop}
\eq
%where $\lambda_{\ell}$ is the lepton Yukawa coupling.
The first term in the numerator arises from the soft scalar tri-linear
$A$ term mixing left- and right-handed sleptons through the down-type
Higgs condensate, while the second term arises from a superpotential
cross term between the Higgsino mass parameter and lepton Yukawa coupling
through the up-type Higgs condensate.
Through second order in mixing, the chirality violating neutralino propagator
of this diagram is dominated by the Bino component.
In Weyl notation this propagator is
\beq
\langle \bino \bino \rangle \simeq
% { i~m_{\wino}^* \over p^2 - |m_2|^2} ~
% { (-i) g_1 v_u^* \over \sqrt{2} } ~
% { i~\mu^* \over p^2 - |\mu|^2}
%{ i \pslash(-i g_1 v_d / \sqrt{2} ) i \pslashb ~+~
  {i m_{\bino}^*  \over
 (p^2 - |m_{\bino}|^2) }
\label{BBprop}
\eq
The dipole operator phase arising from the first term
in slepton propagator (\ref{sleptonprop}) proportional to the down-type
Higgs condensate along with the phase of the Bino propagator (\ref{BBprop})
is ${\rm Arg}({\cal D}_{\ell})={\rm Arg}(\lambda_{\ell} A v_d m_{\bino}^*)$.
The physical phase ${\rm Arg}({\cal D}_{\ell} m_{\ell}^*)$ from these terms
is therefore given by the basis independent combination of phases \cite{phase}
\beq
\varphi \simeq {\rm Arg} \left(A m^*_{\bino} \right)
\label{bdphase}
\eq
With slepton proportionality
the magnitude of this term is however suppressed with respect to the
second term in the slepton propagator for moderate to large
$\tan \beta$ since it is proportional to the down-type Higgs condensate.
The dipole operator phase arising from the second term in the slepton
propagator (\ref{sleptonprop}) proportional to the up-type Higgs condensate
along with the phase of the Bino propagator (\ref{BBprop})
is ${\rm Arg}({\cal D}_{\ell})={\rm Arg}(\lambda_{\ell} \mu^* v_u^* m_{\bino}^*)$.
Anti-alignment of the relative phase of the up- and down-type Higgs condensates
with the Higgs up--Higgs down soft mass parameter,
${\rm Arg}(v_u v_d) = - {\rm Arg}(m_{ud}^2)$, then implies that the
physical phase ${\rm Arg}({\cal D}_{\ell} m_{\ell}^*)$ is given
by the basis independent combination of phases (\ref{EDMphase}).
So under the assumption of slepton universality, proportionality
and gaugino unification,
{\it all} the $\tan \beta$ enhanced electromagnetic dipole
operator diagrams have the same phase to leading order in
Higgsino-gaugino mixing.
The phase (\ref{EDMphase}) therefore dominates the phase appearing in the
electron EDM
for moderate to large $\tan \beta$ over most of parameter space.

It might have been possible in principle for the phases
among various contributions to the dipole operator
to have accidentally approximately canceled \cite{cancel}.
This could in principle occur for small $\tan \beta$
by a cancelation between the phases (\ref{EDMphase}) and
(\ref{bdphase}).
But the Bino diagram proportional to the phase (\ref{bdphase})
is parametrically suppressed by ratios of gauge couplings
compared with the dominant chargino diagram.
Cancelation would only occur if the ratio of
the phases (\ref{EDMphase}) and (\ref{bdphase})
just happens to be nearly equal in magnitude and opposite
in sign to the ratio of the chargino to Bino contributions.
Cancelations might also in principle occur in the region of
parameter space with large Higgsino-gaugino mixing.
However, such fortuitous
cancelations depend on accidental details of the superpartner spectrum,
are not enforced by any symmetry, and occur
only over very narrow slivers of parameter space \cite{cnote}.
Outside of these narrow regions of parameter space the electron
EDM can therefore be considered to bound the phase (\ref{EDMphase})
rather directly for moderate to large $\tan \beta$ under the assumption
of slepton universality and proportionality.

Slepton flavor violation can in principle lead to
violations of the proportionality relation
(\ref{relation}). %as discussed in section {\ref{opsection}),
Left--left and right--right sflavor violation in the slepton propagators
allows flavor conserving left--right mass squared insertions proportional 
to both $m_{\mu}$ and $m_{\tau}$ in the electron 
electromagnetic dipole operator as illustrated in (\ref{desflavor}).
%The left--left and right--right sflavor violating mass squared 
%insertions which give rise to the sflavor violation in this case can
This sflavor violation can 
introduce important additional sources for the
physical phase appearing in the electron EDM.
The intermediate stau contribution to the electron EDM 
proportional to $m_{\tau}$ through a left--right mixing 
arises from the second neutralino diagram of Fig. \ref{figv}
with both left--left and right--right selectron--stau
flavor violating mass squared insertions.
The ratio of the stau--Bino contribution to the electron EDM
to the flavor conserving contribution from the selectron--chargino 
first diagram of Fig. \ref{figv} is 
\beq
{ \desusystau \over \desusyse } \simeq
 \left({g_1^2 \over  3 g_2^2} {\mtau \over m_{e} }
  {m_{\bino} \over m_{\wino} } \right)~
  |\dacLL \dcaRR| ~ {h_0 \over f_{+}} h^{\prime \prime}_{0,LR}
  ~ {\sin (\varphi + \phac ) \over \sin \varphi }
\label{seratio}
\eeq
where $\varphi$ is the relative phase between the
flavor conserving contribution to the dipole operator
and electron mass (\ref{relphase}) and 
\beq
\phac = {\rm Arg}\left( \dacLL \dcaRR \right)
\eq
is the phase of the left--left times right--right selectron--stau
mass squared mixing,
and where strict gaugino unification, ${\rm Arg}(m_{\bino}) =
{\rm Arg}(m_{\wino})$, has been assumed.
The functions $f_+$ and $h_0$
are loop functions defined in the appendix
for the chargino first diagram of
Fig. \ref{figv} and the neutralino second diagram
of Fig. \ref{figv}, and normalized to unity for equal superpartner
masses.
The dimensionless derivative function
$h^{\prime \prime}_{0,LR}$ defined in (\ref{dlogfcn})
represents the modification of the loop function induced by
the two sflavor violating mixing insertions such that
$h_0 h^{\prime \prime}_{0,LR}$ is the loop function for the stau
contribution $\desusystau$.

The importance of the sflavor violating stau contribution 
to the electron EDM depends on the magnitude and 
phases of the left--left and right--right selectron--stau 
mass squared mixings. 
With gaugino unification, the first term in parenthesis 
on the right hand side of the ratio (\ref{seratio}) is 
$(m_{\tau} / 3 m_e) \tan^4 \theta_w \simeq 100$. 
%If the current discrepancy between $\amuexp$ and 
%$\amususy$ \cite{bnl} is interpreted as arising from 
%supersymmetry the current limits on the 
The most stringent possible limits on the 
magnitude of left--left and right--right selectron--stau 
sflavor violation arising from the limits on 
$\tau \rightarrow e \gamma$ radiative decay, 
and consistent with the current experimental results 
for the muon anomalous magnetic moment, are presented in 
section \ref{decaysection}. 
The bounds derived there imply that at best  
$ |\dacLL \dcaRR| \lsim 10^{-1}$
up to ratios of model dependent loop functions. 
Since the sflavor violating phases are unconstrained, 
the $m_{\tau}$ enhanced sflavor violating stau contribution
to the electron EDM could clearly dominate the flavor 
conserving contribution. 
So unlike the muon anomalous magnetic moment discussed 
in section \ref{muonsection}, the electron EDM can 
potentially receive significant contributions from 
sflavor violation. 

The muon anomalous magnetic moment 
is likely to be dominated by sflavor conserving contributions
for any $\amususy$ which will be accessible to the 
ultimate sensitivity of the Brookhaven muon $g-2$ 
experiment \cite{bnl2}, 
as discussed in section \ref{muonsection}.
The sflavor conserving supersymmetric
contribution $\amususy$ may then still be used to characterize
the magnitude of the sflavor violating stau contribution 
to the electron EDM in terms of sflavor violation and ratios
of loop functions. 
From the ratio (\ref{seratio}) and the relation between the 
flavor conserving contribution to the electron EDM and 
$\amususy$ given in (\ref{edmrelation}),
the sflavor violating stau contribution may be written 
\ba
 \desusystau & \simeq &  - { g_1^2 \over 6 g_2^2}
 {m_{\tau} \over m_{\mu}^2 }
 {m_{\bino} \over m_{\wino} }
   ~\amususy ~
 |\dacLL \dcaRR| ~ {h_0 \over f_{+}} h^{\prime \prime}_{0,LR}~\times
   \nonumber \\
 & & ~~~~
 \left( \tan \varphi \cos \phac + \sin \phac \right)
   ~~e~{\rm cm} \nonumber \\
  & \simeq &
  -4.9 \times 10^{-14}
  ~\amususy~
  |\dacLL \dcaRR| ~ {h_0 \over f_{+}} h^{\prime \prime}_{0,LR}~\times
    \nonumber \\
  & & ~~~~
 \left( \tan \varphi \cos \phac + \sin \phac \right)
  ~~e~{\rm cm}
\label{edmstaurelation}
\ea
If the current discrepancy
between $\amuexp$ and $\amusm$ \cite{bnl} is
interpreted as arising from supersymmetry,
$\amususy \sim 42 \times 10^{-10}$,
the current bound on the electron EDM of
$|d_e| < 4 \times 10^{-27}~e~{\rm cm}$ obtained from
$^{205}$Tl \cite{edmbound} along with the relation
(\ref{edmstaurelation}) can be used to obtain a bound on the
imaginary part of the product of the left--left times right--right
selectron--stau mass squared mixings
$$
|\dacLL \dcaRR| |\sin \phac | \lsim
   2 \times 10^{-5} ~
   \left( {h_0 \over f_{+}} h^{\prime \prime}_{0,LR} \right)^{-1}
$$
where possible cancelation with the flavor conserving contribution
has been ignored.
If the sflavor violating 
phase are large, the electron EDM apparently provides a 
more stringent bound on the the product 
left--left times right--right selectron--stau 
mixing than that obtained from $\tau \rightarrow e \gamma$ 
decay discussed in section \ref{decaysection}.
Alternatively, an observation of $\tau \rightarrow
e \gamma$ close to the current 
experimental limit  combined 
with the
electron EDM constraint would yield a very strong 
direct bound on this product of sflavor violating phases.

%For ${\cal O}(1)$ phases in the product of the
%left--left times right--right
%selectron--stau mass squared mixings the electron EDM
%provides a stronger constraint on the flavor mixing parameters
%then obtained from the $\tau \rightarrow e \gamma$ decay.
%Stronger limits may be obtained from $\mu \rightarrow e
%\gamma$ if both 1-3 and 2-3 mixing is present. Within
%a model of flavor this translates
%into a direct bound on 1-3 mixing, but as this is model--dependent
%this information is not used.

%**************************

%The flavor violating contributions to the electron EDM
%may occur from either stau loops or smuon loops, and
%can be enhanced relative to the universal contribution by
%an amount $m_{f=\tau,\mu}/m_e$. These diagrams are
%the same $LR$ mass insertion diagrams discussed above,
%but now with either 1--3 or 1--2 mixing.

The electron EDM can also receive analogous sflavor violating contributions
proportional to $m_{\mu}$ through sflavor conserving 
left--right smuon mixing
in combination with 
both left--left and right--right selectron--smuon mixing, 
as illustrated in (\ref{desflavor}).
The ratio of this contribution to the sflavor conserving chargino 
contribution 
is identical to (\ref{seratio}) with $m_{\tau}$ replaced by $m_{\mu}$,
and stau mixing replaced by smuon mixing. 
This contribution should be smaller than the possible stau contribution 
discussed above for two reasons. 
First, the smuon contribution is proportional to $m_{\mu}$ rather
than $m_{\tau}$. 
Second, the bounds on selectron--smuon mixing 
%from 
%$\mu \rightarrow e \gamma$ radiative decay are much more stringent 
%than the analgous tau decay. 
are much more stringent than those on selectron--stau mixing. 
The most stringent possible limits derived is section \ref{decaysection}
on selectron--smuon sflavor violation arising from limits
on $\mu \rightarrow e \gamma$ radiative decay, and consistent with
the current experimental results for the muon anomalous magnetic 
moment, imply that at best 
$|\dabLL \dbaRR | \lsim 10^{-7}$ up to ratios of model dependent 
loop functions. 
In this case the smuon contribution to the electron EDM 
is smaller than the sflavor conserving selectron--chargino
contribution unless the the slfavor conserving phase
is smaller than the sflavor violating phase
by a factor of roughly $10^{-6}$.

\section{Radiative Flavor Changing Lepton Decays}
\label{decaysection}

Non--trivial flavor structure of the electromagnetic dipole
operators gives rise to radiative flavor changing
fermion decays.
There are two possible chiral structures for such
transition dipole moments.
For example, for $\mu$--$e$ transitions the operators are
\beq
- {1 \over 2} {\cal D}_{L e \mu }~ \overline{e}_L \sigma^{\mu \nu}
 \mu_R F_{\mu \nu}
 - {1 \over 2} {\cal D}_{L e \mu }^* ~\overline{\mu}_R \sigma^{\mu \nu}
 e_L F_{\mu \nu}
\label{tranR}
\eeq
\beq
- {1 \over 2} {\cal D}_{R e \mu }~ \overline{e}_R \sigma^{\mu \nu}
 \mu_L F_{\mu \nu}
 - {1 \over 2} {\cal D}_{R e \mu }^* ~\overline{\mu}_L \sigma^{\mu \nu}
 e_R F_{\mu \nu}
\label{tranL}
\eeq
where the subscript $L$ or $R$ refers to the chirality of the
lighter final state fermion, and likewise for $\tau$--$\mu$ and
$\tau$--$e$ operators.
The lepton radiative flavor changing decay rates arising
from transition operators of the form
(\ref{tranR}) and (\ref{tranL}) are
\beq
\Gamma( \ell_i \rightarrow \ell_j \gamma) =
{ ( |{\cal D}_{L ij}|^2 + |{\cal D}_{Rij}|^2) ~m_{\ell_i}^3
     \over 16 \pi}
\label{raddecay}
\eeq
where the left- and right-handed operators do not
interfere %with vanishing final state masses \cite{helicity}.
up to corrections of order $m_j^2/m_i^2$ \cite{helicity}.

The supersymmetric Standard Model allows for the possibility
of individual lepton flavor violation in the left- and
right-handed slepton
soft mass squared matrices and in the scalar tri-linear
soft $A$-terms mixing left- and right-handed sleptons.
These flavor violations appear in the slepton
propagators after electroweak symmetry breaking.
The supersymmetric diagrams which contribute to the
transition dipole operators are just those
of section \ref{opsection} with the inclusion of
slepton flavor violating propagators.
If the flavor violation in these propagators is small,
it may be represented by left--left, right--right
and left--right flavor violating mass squared
insertions.
In this case the magnitude of the flavor
violating operators may be related to that of the
flavor conserving operators in terms the small flavor
violation.
Under the various assumptions detailed below, the
muon anomalous magnetic moment may then
be related to the decay rates of radiative flavor changing decays
$\ell_i \rightarrow \ell_j \gamma$ in terms
of flavor violating mass squared insertions.

Consider first the assumption of
approximate slepton universality and approximate
proportionality which we define as sflavor violation
which is small enough so as not to modify the relation
(\ref{relation})
that flavor conserving dipole operators are in proportion
to the fermion masses.
In this case the leading effects come
from single insertions of flavor violating mass squared insertions.
Possible modifications of approximate  universality and
proportionality arising from sflavor
violation are considered separately below.
As discussed in section \ref{opsection}
the chirality violating chargino--sneutrino first diagram of
Fig. \ref{figv} generally gives the dominant contribution
to the flavor conserving dipole operators for moderate
to large $\tan \beta$.
In this case, ignoring possible
cancelations with sub-dominant diagrams discussed below,
the supersymmetric
contributions to the transition
dipole moment for $\ell_i \rightarrow \ell_j \gamma$
can be related to the flavor conserving
muon dipole moment in terms of sneutrino
flavor violating left--left mass squared insertions
in the chargino-sneutrino diagram
as
\beq
{\cal D}_{L i j } \supset {m_i \over m_{\mu}}~
{\cal D}_{\mu} ~ \dijLL ~ f_{+,L}^{\prime}
%{\delta m^2_{\selectron_L \smuon_L}
%  \over m^2_{\slepton_L} }
\label{dij}
\eeq
where the flavor conserving
${\cal D}_i \simeq (m_i / m_{\mu}) {\cal D}_{\mu}$ is assumed
to be dominated by chargino-sneutrino diagram, and
$\dijLL$ is the dimensionless sflavor violating
left--left mixing defined in terms of the
left--left mixing mass squared matrix in (\ref{deltadef}).
The dimensionless derivative function
\beq
f_{+,L}^{\prime} \equiv \frac{ m^2_{\sneutrino} }{f_+}
\frac{\partial f_+}{\partial m^2_{\sneutrino} }  =
\frac{ \partial \ln f_+}{\partial \ln m^2_{\sneutrino} }
\label{dlijrelationLL}
\eeq
represents the modification of the loop function induced
by the left--left sflavor violating insertion
on the sneutrino propagator such that
$f_+f_{+,L}^{\prime}$ is the loop function for the transition dipole
operator with $f_+=f_+(m^2_{\sneutrino}, \mwino^2, \mu^2)$
the loop function of the dominant chirality violating
chargino first diagram of Fig. \ref{figv}.
This order one function depends on details of the superpartner mass
spectrum.
For equal superpartner masses $f_{+,L}^{\prime}=-0.4$, while for
$0.1< m^2_{\sneutrino}/\mwino^2 < 10$
with $\mwino=\mu$ it varies in the range
$-0.11 < f_{+,L}^{\prime}< -0.75$.
Aside from this slight model dependence from
modification of the loop function,
the transition dipole moment induced by left--left sflavor violation
is related rather directly to the muon dipole moment
by (\ref{dlijrelationLL})
for moderate to large $\tan \beta$ and assuming approximate
universality.

Transition dipole moments can also receive contributions from
left--right and right--right sflavor violation through diagrams
which are sub-dominant in the flavor conserving dipole moments.
The importance of these diagrams depends on the
relative magnitude of the underlying sflavor violations.
Left--right sflavor violation contributes to transition dipole
operators through the chirality violating neutralino
second diagram of Fig. \ref{figv}.
The left--right slepton mixing insertion
represented by the dot in the second diagram Fig. \ref{figv}
in this case is flavor violating.
Since this diagram does not involve an explicit Yukawa coupling
in the neutralino coupling, it contributes to both
chiralities of transition operators (\ref{tranR}) and (\ref{tranL})
\beq
{\cal D}_{L i j }~, ~{\cal D}_{R i j } \supset {m_i \over m_{\mu}}~
{\cal D}_{\mu} ~
{g_1^2 \over 3g_2^2} {m_{\bino} \over m_{\wino}}
{h_0 \over f_+} ~
{m_{\slepton_L}m_{\slepton_R} \over m_i \mu \tan \beta}
  ~\dijRL
\eeq
The factor $(g_1^2 \mbino / g_2^2 \mwino)(h_0/f_+)$ accounts for the
difference in loop function and parametric dependence of the couplings and
gaugino mass insertions and of the subdominant neutralino
second diagram of Fig. \ref{figv} compared with the dominant
chargino first diagram.  The loop functions are normalized
to unity for equal superpartner masses.
The factor $m_{\slepton_L}m_{\slepton_R}/(m_i \mu \tan \beta)$
accounts for the difference in parametric dependence of the
flavor conserving and violating left--right mass squared
insertions.

Right--right sflavor violation contributes to transition dipole
operators at lowest order through two diagrams.
The first is the chirality violating neutralino
first diagram of Fig. \ref{figv} with a right--right mass squared
mixing insertion on the slepton propagator.
The second is the chirality violating neutralino
second diagram of Fig. \ref{figv} with again a right--right mass
squared insertion on the right handed slepton propagator.
Both these diagrams have the same parametric dependence
on couplings and chirality violating mass insertions
on the neutralino propagators
\beq
{\cal D}_{Ri j } \supset {m_i \over m_{\mu}}~
{\cal D}_{\mu} ~ {g_1^2 \over 3g_2^2} {m_{\bino} \over m_{\wino}}
\left(
{-f_0 \over f_+}f_{0,R}^{\prime}
+{h_0 \over f_+}h_{0,R}^{\prime} \right) \dijRR
\eeq
where $\dijRR$ is the dimensionless
slepton sflavor violating right-right mixing.

In order to display the relation between the transition
dipole moments, the muon anomalous magnetic
moment which determines the overall scale for the
dipole moments, and sflavor violating mass squared insertions,
it is convenient to define the transition dipole operator
coefficients in terms of the scaled flavor conserving  dipole
operator coefficient times dimensionless flavor violating
transition elements
$$
{\cal D}_{Lij} \simeq - e {m_i \over 2 \mmu ^2} \amususy~
\eLij
$$
\beq
{\cal D}_{Rij} \simeq - e {m_i \over 2 \mmu ^2} \amususy~
\eRij
\eeq
The leading sflavor violating contributions discussed above
then give
\beq
\eLij \simeq  \dijLL ~ f_{+,L}^{\prime} +
{g_1^2 \over 3g_2^2} {m_{\bino} \over m_{\wino}}
{h_0 \over f_+} ~
{m_{\slepton_L}m_{\slepton_R} \over m_i \mu \tan \beta}
  ~\dijRL
  \label{telL}
\eq
\beq
\eRij \simeq
{ g_1^2 \over 3g_2^2} {m_{\bino} \over m_{\wino}}
\left[ \left(
{-f_0 \over f_+}f_{0,R}^{\prime}
+{h_0 \over f_+}h_{0,R}^{\prime} \right) \dijRR
+ {h_0 \over f_+} ~
{m_{\slepton_L}m_{\slepton_R} \over m_i \mu \tan \beta}
  ~\dijRL
  \right]
  \label{telR}
\eq
%so that the $\eLij$ and $\eRij$ depend only on the
%ratio of couplings, loop functions, and the dimensionless sflavor
%violating mixings.
%5This electron--muon transition dipole moment
%ives rise to
%muon radiative branching ratio of
Supersymmetric contributions
%from the
%transition dipole moment coefficient (\ref{dij}) to the
to the branching
branching ratios for radiative flavor changing decays
$\ell_i \rightarrow \ell_j \gamma$ relevant for muon and tau
decays are then
%(\ref{taurelation}) with (\ref{raddecay}) gives
\ba
{\rm Br}(\ell_i \rightarrow  \ell_j \gamma) & \simeq &
{ 12 \pi^3 \alpha~  ({\amususy})^2  \over   {G_F^2}  ~m^4 _{\mu} }
%      {\delta m^2_{\selectron_L \smuon_L} \over m^2_{\slepton_L} }
 ~ \left( |\eLij |^2 +  |\eRij |^2 \right)~
  {\rm Br}(\ell_i \rightarrow \ell_j \overline{\nu}_{\ell_j} \nu_{j})
\nonumber \\
%\ba
%{\rm Br}(\ell_i \rightarrow  \ell_j \gamma) & \simeq &
%12 \pi^3 \alpha
%\left( { a^{\rm SUSY}_{\mu}  ~\zeta_+
%     |(\delta^{\ell}_{ij})_{LL}| \over G_F ~m_{\mu}^2 }
%%      {\delta m^2_{\selectron_L \smuon_L} \over m^2_{\slepton_L} }
%\right)^2  ~
%  {\rm Br}(\ell_i \rightarrow \ell_j \overline{\nu}_{\ell_j} \nu_{j})
%\nonumber \\
 & & \nonumber \\
  & \simeq & 1.85 \times 10^{14}
   (\amususy)^2
%   {\delta m^2_{\selectron_L \smuon_L} \over m^2_{\slepton_L}} \right)^2
%|(\delta^{\ell}_{ij})_{LL}| \right)^2
  \left( |\eLij |^2 + |\eRij |^2 \right)
  {\rm Br}(\ell_i \rightarrow \ell_j \overline{\nu}_{\ell_j} \nu_{j})~~
%\left(\frac{a^{\rm SUSY}_{\mu} |(\delta^{l}_{12})_{LL}| h}{G_F m^2_{\mu}}
%\right)^2 .
\label{leptondecay}
\ea
where $\alpha$ is the fine structure constant,
the muon weak decay rate is
$\Gamma(\mu \rightarrow e \overline{\nu}_e \nu_{\mu})
 = G^2_F m^5_{\mu} / (192 \pi^3)$,
${\rm Br}(\mu \rightarrow e \overline{\nu}_{e} \nu_{\mu})
 \simeq 1$, and
${\rm Br}(\tau \rightarrow \ell_j \overline{\nu}_{\ell_j} \nu_{\tau})
 \simeq 0.175$.
Note that the $m_i$ dependence of the transition dipole
coefficient (\ref{dij}) cancels with the $m_i$ dependence
of the radiative decay rate (\ref{raddecay}) within the branching ratio
(\ref{leptondecay}).

Left--left and right--right sflavor violating slepton mass squared
mixings appear in the dimensionless flavor violating transition
elements (\ref{telL}) and (\ref{telR}) in proportion to ratios of
coupling constants, Bino to Wino masses which may be related
to coupling constants under the assumption of gaugino unification
$m_{\bino} / m_{\wino} = g_1^2 / g_2^2$, and ratios and
derivatives of loop functions.
Since, as discussed above, the ratios and derivatives
of loop functions do not depend to drastically on the superpartner
spectrum, left--left and right--right sflavor violating contributions
to radiative flavor changing decays
can then be related to the muon anomalous magnetic moment
through the branching ratios (\ref{leptondecay})
in a fairly model independent manner.
In contrast left--right sflavor violating mass squared mixings appear
in the transition elements (\ref{telL}) and (\ref{telR})
in proportion to slepton masses and $\tan \beta$.
The relation between left--right mixings, the muon anomalous magnetic
moment, and flavor changing
branching ratios (\ref{leptondecay}) is therefore model
dependent.
Therefore only the fairly model independent relations which
can be extracted for left--left and right--right sflavor violating
mass squared
mixings are presented below.

%This relation between
%${\rm Br}(\ell_i \rightarrow  \ell_j \gamma)$, $\amsusy$,
%and is good for moderate to large .....
%\dijLL ~ f_{+,L}^{\prime}
%{\delta m^2_{\selectron_L \smuon_L}
%  \over m^2_{\slepton_L} }
%\label{dij}
%\eeq

If the current discrepancy \cite{bnl}
between $\amuexp$ and $\amusm$ is
interpreted as arising from supersymmetry,
$\amususy \sim 43 \times 10^{-10}$,
the current bound on $\mu \rightarrow e \gamma$ of
${\rm Br}(\mu \rightarrow e \gamma) < 1.2 \times 10^{-11}$
\cite{muebound} along with the relation (\ref{leptondecay})
can be used to obtain a bound on the smuon--selectron
transition dipole coefficients of
$$
%{\delta m^2_{\selectron_L \smuon_L} \over m^2_{\slepton_L}}
\sqrt{|\eLba|^2 + |\eRba|^2}
   \lsim 6 \times 10^{-5} ~
   %|\zeta|^{-1}   % / \sqrt{\zeta}
   \label{muemixingbound}
$$
In the absence of cancelations among the various contributions,
%and assuming a single contribution dominates this bound
this bound on the dipole coefficients may be used to bound the
individual dimensionless sflavor violating mixings.
Assuming a single contribution dominates then yields
bounds on left--left and right--right selectron--smuon dimensionless
mass squared mixing of
$$
|\dabLL| \lsim 6 \times 10^{-5} ~(f_{+,L}^{\prime})^{-1}
$$
$$
|\dabRR| \lsim 2 \times 10^{-3} ~
 \left(
{-f_0 \over f_+}f_{0,R}^{\prime}
+{h_0 \over f_+}h_{0,R}^{\prime} \right)^{-1}
$$
%$$
%|\dabLR| \lsim X \times 10^{-X} ~ ....
%$$
where gaugino unification
$m_{\bino} / m_{\wino} = g_1^2 / g_2^2 = \tan^2 \theta$ has been
assumed.
%The bound on $\dabLL$ is the most model independent

The bound obtained above on $\dabLL$ is stronger by 
roughly a factor of 50--100
than a previously quoted bound \cite{masiero}. 
The difference arises from a number of factors. 
In \cite{masiero}
only the chirality conserving
photino diagram is considered, which
is suppressed compared to the dominant 
chargino diagram by factors of gauge
couplings, $\tan \beta$, and for equal superpartner masses, a
smaller loop function.
In addition there has also been an
improvement in the experimental sensitivity to $\mu \rightarrow e \gamma$ 
radiative decays \cite{muebound}.
All these factors combine to provide a stronger constraint.

In analogy the the bounds derived above, the
current bounds on flavor changing $\tau$ radiative decays of
${\rm Br}(\tau \rightarrow \mu \gamma) < 1.1 \times
  10^{-6}$ \cite{taumubound} and
${\rm Br}(\tau \rightarrow e \gamma) < 2.7 \times
  10^{-6}$ \cite{tauebound}
along with the relation (\ref{leptondecay}) can be used to obtain
bounds on smuon--stau and selectron--stau dipole coefficients of
$$
\sqrt{|\eLcb|^2 + |\eRcb|^2}
   \lsim 4 \times 10^{-2}
$$
$$
\sqrt{|\eLca|^2 + |\eRca|^2}
   \lsim 7 \times 10^{-2}
$$
Assuming a single contribution dominates then yields the
bounds on the left--left smuon--stau and selectron--stau
dimensionless mass squared mixings of
$$
|\dcbLL|
   \lsim 4 \times 10^{-2} ~(f_{+,L}^{\prime})^{-1}
%   \label{taumumixingbound}
$$
$$
|\dcaLL|
   \lsim 7 \times 10^{-2} ~(f_{+,L}^{\prime})^{-1}
%   \label{taumumixingbound}
$$
where again gaugino unification has been assumed.
Bounds on the right--right mixings are not significant
since these insertions appear in subdominant diagrams
$$
|\dcbRR|
   \lsim 1.3 ~
 \left(
{-f_0 \over f_+}f_{0,R}^{\prime}
+{h_0 \over f_+}h_{0,R}^{\prime} \right)^{-1}
   \label{tauemixingbound}
$$
$$
|\dcaRR|
   \lsim 2.3 ~
 \left(
{-f_0 \over f_+}f_{0,R}^{\prime}
+{h_0 \over f_+}h_{0,R}^{\prime} \right)^{-1}
   \label{tauemixingbound}
$$
All the bounds given above are good for moderate to large
$\tan \beta$ for which the first chargino diagram of Fig. \ref{figv}
generally gives the dominant contribution.
The only model dependence appears in the ratios and logarithmic
derivatives of loop functions as indicated.
If the current discrepancy \cite{bnl} between
$\amuexp$ and $\amusm$ is
interpreted as an upper limit on supersymmetric contributions,
$\amususy \lsim 43 \times 10^{-10}$,
then the bounds given above may be interpreted as the
most stringent possible bounds consistent with the muon
anomalous magnetic moment.

%The supersymmetric contribution to $\mu \to e \gamma$
%may then be related to the supersymmetric contribution to
%the muon anomalous magnetic moment by
%\ba
%{\rm Br}(\mu \rightarrow  e \gamma) & \simeq &
%12 \pi^3 \alpha
%\left( { a^{\rm SUSY}_{\mu}  ~\zeta ~
%     |(\delta^{\ell}_{12})_{LL}|\over G_F~ m_{\mu}^2 }
%%      {\delta m^2_{\selectron_L \smuon_L} \over m^2_{\slepton_L} }
%\right)^2  \nonumber \\
% & & \nonumber \\
%  & \simeq & 1.85 \times 10^{14}~ \left(  a^{\rm SUSY}_{\mu}  ~\zeta~
%%   {\delta m^2_{\selectron_L \smuon_L} \over m^2_{\slepton_L}} \right)^2
%|(\delta^{\ell}_{12})_{LL}| \right)^2
%%\left(\frac{a^{\rm SUSY}_{\mu} |(\delta^{l}_{12})_{LL}| h}{G_F
%m^2_{\mu}}
%%\right)^2 .
%\label{muebr}
%\ea

%$$
%(\delta^{\ell}_{12})_{LL} \equiv
%   {\delta m^2_{\selectron_L \smuon_L} \over m^2_{\slepton_L} }
%$$

%The radiative flavor changing decay $b \rightarrow s \gamma$
%also arises from a transition dipole operator and is
%sensitive to supersymmetric contributions proportional
%to the top Yukawa squared from virtual stops and charginos.
%However, extension of a model independent relation between
%$a_{\mu}^{\rm SUSY}$ and
%supersymmetric contributions to $b \rightarrow s \gamma$ is
%complicated first by model dependence of the ratio of
%squark to slepton masses, and second by the significant
%charged Higgs contribution to $b \rightarrow s \gamma$.

% ---------------------------------------------------

\section{Conclusions}

Supersymmetry can give many interesting signals which may
be observable in low energy processes.
In this paper we have illuminated the relation between the
muon anomalous magnetic moment, the electron EDM,
and the lepton flavor violating radiative decays,
$\mu \rightarrow e \gamma$,
$\tau \rightarrow \mu \gamma$ and
$\tau \rightarrow e \gamma$.
A bound(measurement) of the non-Standard Model contributions
to the muon anomalous magnetic moment can
bound(determine) the overall scale of the dipole operators
which contribute to these processes in a manner which is
fairly insensitive to details of the superpartner
mass spectrum.
In this way the electron EDM and $\ell_i \rightarrow \ell_j \gamma$
decays can be related to small supersymmetric
violations of time-reversal and slepton flavor respectively
in a fairly model independent manner.

The Brookhaven muon $g-2$ experiment may eventually reach a sensitivity
of $\Delta a_{\mu}^{\rm exp} \sim 4  \times 10^{-10}$ \cite{bnl2}.
If this sensitivity is achieved
and the measured value is in agreement with an improved
determination of the Standard Model hadronic vacuum
polarization and light by light scattering contributions
then the bound on $a_{\mu}^{\rm SUSY}$ would improve by
approximately an order of magnitude.
Such an agreement with the Standard Model
would imply concomitantly heavier superpartners and
weaken the bound given above on the phase of the dipole
operator %\ref{phasebound}
coming from the present
$^{205}$Tl EDM experiment \cite{edmbound} by an order of magnitude,
and weaken the bounds given above on the slepton mixing
amplitudes
coming from the present $\ell_i \to \ell_j \gamma$
experiments \cite{muebound,taumubound,tauebound}
by a factor of roughly three.
However, future EDM experiments in atomic traps may improve
the sensitivity to the phase of the dipole operator by
two to three orders of magnitude \cite{chu}, and
a future experiment sensitive to
${\rm Br}(\mu \to e \gamma) \gsim 2 \times 10^{-14}$ \cite{newmue}
would improve the sensitivity to the selectron--smuon mixing amplitude
by a factor of roughly 25.

\vspace{.15in}

The work of M.G. was supported by US Department of Energy.
The work of S.T. was supported by
the US National Science Foundation under grant PHY98-70115,
the Alfred P. Sloan Foundation, and Stanford
University through the Frederick E.~Terman Fellowship.

% --------------------------------------------------------------

\appendix{}
\section{Supersymmetric Dipole Moments}

%The supersymmetric one-loop contributions...........
%large $\tan \beta$ and to first order in the gaugino-Higgsino
%mixing.
Supersymmetric contributions to lepton electromagnetic dipole
operators which couple to the up-type Higgs condensate are
enhanced by a factor $\tan \beta$ with respect to couplings
to the down-type Higgs condensate.
As discussed in section \ref{opsection} the contributions which
are enhanced by this factor include a subset of the
chirality violating diagrams of Fig. \ref{figv}.
In this appendix the expressions for these $\tan \beta$
enhanced contributions are presented.
Gaugino--Higgsino mixing is treated perturbatively to first order
as an insertion of a up-type Higgs condensate.
Slepton proportionality is assumed in which the left-right
mass squared mixing is given by $m_{\ell} \mu \tan \beta$
in the large $\tan \beta$ limit.

The supersymmetric contributions
to ${\cal D}_{f}$ may be expressed in terms of the
dimensionless loop functions %$I_N$ and $J_N$ functions, where
\beq
J_N(x_1,x_2,\cdots x_N) \equiv \int^{\infty} _{0}
dy ~y^2 \prod_{i=1} ^{N} \frac{1}{y+x_i}
\eeq
\beq
 I_N(x_1,x_2,\cdots x_N) \equiv \int^{\infty} _{0}
dy ~y  \prod_{i=1} ^{N} \frac{1}{y+x_i}
\eeq
%This follows the notation of \cite{moroi}.
In terms of these functions, the
chargino with sneutrino, Wino neutralino with left-handed slepton,
Bino neutralino with left-handed slepton, Bino
neutralino with right-handed slepton, and Bino
neutralino with left-right mass squared insertion
diagrams of Fig. \ref{figv} give, respectively,
\ba
\langle \tilde{W}^{+} \tilde{H}^- _d  \rangle
\hbox{  :  }
-2 {\cal D}^{\chi^{+}}_{f} &=& \frac{e
g^2_2 \tan \beta}{32 \pi^2}
\frac{m_{f}}{\tilde{m}^4 _H} \mu m_{\tilde{W}}
\left(4 J_5(\frac{m^2 _{\tilde{\nu}_L}}{\tilde{m}^2 _H},
\frac{\mu^2}{\tilde{m}^2_H}, \frac{m^2_{\tilde{W}}}{\tilde{m}^2_H},
 \frac{\mu^2}{\tilde{m}^2_H},\frac{\mu^2}{\tilde{m}^2_H}) \right.
\nonumber \\
& &\left.
+4 J_5(\frac{m^2 _{\tilde{\nu}_L}}{\tilde{m}^2 _H},
\frac{\mu^2}{\tilde{m}^2_H}, \frac{m^2_{\tilde{W}}}{\tilde{m}^2_H},
   \frac{m^2_{\tilde{W}}}{\tilde{m}^2_H},
\frac{m^2_{\tilde{W}}}{\tilde{m}^2_H} )
\right. \nonumber \\
& & \left.
+4 J_5( \frac{m^2 _{\tilde{\nu}_L}}{\tilde{m}^2 _H},
\frac{\mu^2}{\tilde{m}^2_H}, \frac{m^2_{\tilde{W}}}{\tilde{m}^2_H},
  \frac{m^2_{\tilde{W}}}{\tilde{m}^2_H},  \frac{\mu^2}{\tilde{m}^2_H}
) \right)
\\
& \equiv & \frac{e g^2_2 \tan \beta}{32 \pi^2}
\frac{m_{f}}{\tilde{m}^4 _H} \mu m_{\tilde{W}} f_{+} \nonumber \\
%
% ----------------
%
\langle \tilde{W}^0 \tilde{H}^0 _d \rangle
\hbox{  :  } -2 {\cal D}^{\chi_0}_{f} &=&
-\frac{e g^2_2 \tan \beta}{192 \pi^2} \frac{m_{f}}{\tilde{m}^4 _H}
\mu m_{\tilde{W}}^2
\left(12 I_4(\frac{m_{\tilde{W}}}{\tilde{m}^2_H},
  \frac{\mu^2}{\tilde{m}^2_H},\frac{m^2 _{\tilde{f}_L}}{\tilde{m}^2 _H},
\frac{m^2 _{\tilde{f}_L}}{\tilde{m}^2 _H}   )    \right. \nonumber \\
& & \left.  -12
J_5( \frac{m^2 _{\tilde{f}_L}}{\tilde{m}^2 _H},
\frac{\mu^2}{\tilde{m}^2_H}, \frac{m^2_{\tilde{W}}}{\tilde{m}^2_H},
  \frac{m^2 _{\tilde{f}_L}}{\tilde{m}^2 _H},
\frac{m^2 _{\tilde{f}_L}}{\tilde{m}^2 _H})   \right)
\\
& \equiv & -\frac{e g^2_2 \tan \beta}{192 \pi^2}
\frac{m_{f}}{\tilde{m}^4 _H}
\mu m_{\tilde{W}} f_{\tilde{W}_0 L}
\nonumber \\
%
% --------------
%
\langle \tilde{B} \tilde{H}_d \rangle \hbox{  :  }
-2 {\cal D}^{\chi_0 L} _{f}
 &=&
\frac{e g^2_1 \tan \beta}{192 \pi^2} \frac{m_{f}}{\tilde{m}^4 _H}
\mu m_{\tilde{B}}
\left(12 I_4(\frac{m^2_{\tilde{B}}}{\tilde{m}^2_H},
  \frac{\mu^2}{\tilde{m}^2_H},\frac{m^2 _{\tilde{f}_L}}{\tilde{m}^2 _H},
\frac{m_{\tilde{f}_L}}{\tilde{m}^2 _H}   )   \right. \nonumber \\
& & \left. - 12
J_5( \frac{m^2_{\tilde{f}_L}}{\tilde{m}^2 _H},
\frac{\mu^2}{\tilde{m}^2_H}, \frac{m^2_{\tilde{B}}}{\tilde{m}^2_H},
  \frac{m^2 _{\tilde{f}_L}}{\tilde{m}^2 _H},
\frac{m^2 _{\tilde{f}_L}}{\tilde{m}^2 _H})   \right)
\\
& \equiv & \frac{e g^2_1 \tan \beta}{192 \pi^2} \frac{m_{f}}{\tilde{m}^4 _H}
\mu m_{\tilde{B}} f_{\tilde{B} L} \nonumber \\
%
% --------------
%
\langle \tilde{B} \tilde{H}_d  \rangle \hbox{  :  } -2 {\cal D}
^{\chi_0 R} _{f}
 &=&
  \frac{-2 e g^2_1 \tan \beta}{192 \pi^2} \frac{m_{f}}{\tilde{m}^4
_H}
\mu m _{\tilde{B}}
\left(12 I_4(\frac{m^2_{\tilde{B}}}{\tilde{m}^2_H},
  \frac{\mu^2}{\tilde{m}^2_H},\frac{m^2 _{\tilde{f}_R}}{\tilde{m}^2 _H},
\frac{m^2 _{\tilde{f}_R}}{\tilde{m}^2 _H}   )    \right. \nonumber \\
& & \left.  -12
J_5( \frac{m^2_{\tilde{f}_R}}{\tilde{m}^2 _H},
\frac{\mu^2}{\tilde{m}^2_H}, \frac{m^2_{\tilde{B}}}{\tilde{m}^2_H},
  \frac{m^2 _{\tilde{f}_R}}{\tilde{m}^2 _H},
\frac{m^2 _{\tilde{f}_R}}{\tilde{m}^2 _H})   \right)
\\
& \equiv & \frac{-2 e g^2_1 \tan \beta}{192 \pi^2} \frac{m_{f}}{\tilde{m}^4
_H}
\mu m_{\tilde{B}} f_{\tilde{B} R} \nonumber \\
%
% --------------
%
 \langle \tilde{B} \tilde{B} \rangle \hbox{  :  } -2
{\cal D}^{\chi_0 LR} _{f}
 &=&
\frac{2 e g^2_1 \tan \beta}{192 \pi^2} \frac{m_{f}}{\tilde{m}^4_{H}}
\mu  m_{\tilde{B}}  \left(6 J_5(\frac{m^2_{\tilde{B}}}{\tilde{m}^2_H},
\frac{m^2_{\tilde{B}}}{\tilde{m}^2_H},
\frac{m^2 _{\tilde{f}_L}}{\tilde{m}^2 _H} ,
\frac{m^2 _{\tilde{f}_R}}{\tilde{m}^2 _H},
\frac{m^2 _{\tilde{f}_R}}{\tilde{m}^2 _H}) \right. \nonumber \\
& & \left. +6 J_5(\frac{m^2_{\tilde{B}}}{\tilde{m}^2_H},
\frac{m^2_{\tilde{B}}}{\tilde{m}^2_H},
\frac{m^2 _{\tilde{f}_L}}{\tilde{m}^2 _H},
\frac{m^2 _{\tilde{f}_L}}{\tilde{m}^2 _H},
\frac{m^2 _{\tilde{f}_R}}{\tilde{m}^2 _H}) \right)  \\
& \equiv & \frac{2 e g^2_1 \tan \beta}{192 \pi^2} \frac{m_{f}}
{\tilde{m}^4_{H}}\mu  m_{\tilde{B}} h_0 ~.
\ea
In the text $f_{\tilde{B}R}$ is denoted by $f_0$.
The dependence of the loop functions $f_i$ and $h_0$ on the
mass ratios has been left implicit. The three neutralino
diagrams of the first diagram in Fig. 1 are proportional
to the same loop function evaluated
with different arguments, as is
evident from the above expressions. Also note
that
the factors in the large parentheses or, equivalently,
the loop functions $f_i$ and $h_0$, are normalized to unity for
equal superpartner masses. In particular,
for equal superpartner masses
$J_5(1,1,1,1,1)=1/12$ and $I_4(1,1,1,1)=1/6$.
Thus, for example, for such a
spectrum the chargino
diagram is $\propto (32 \pi^2)^{-1}$. Note that a factor
of $\msusy^{-4}$ has been factored out of the loop
integrals in order to render them dimensionless; this
may be any internal mass but for convenience it is chosen to the
mass of the heaviest sparticle.
The arbitrariness of this factoring follows from the scaling
relations
\beq
J_N(x_1,\cdots, x_N) = x_1^{3-N} J_N(1,\frac{x_2}{x_1}, \cdots, \frac{x_N}{x_1})
\label{scalej}
\eeq
\beq
I_N(x_1,\cdots, x_N) = x_1^{2-N} I_N(1,\frac{x_2}{x_1}, \cdots, \frac{x_N}{x_1})
~.
\label{scalei}
\eeq
%There is a similar expression for $I_N$ but
%with the prefactor changed to $x_1^{2-N}$.
The contributions given above
for the individual
diagrams agree with those found
in \cite{moroi}. There appears to be a difference
in the expressions for the chargino diagram and the
first neutralino diagram of Fig. 2. This
superficial difference is however a result of a
different choice of routing the loop momenta.
A numerical comparison between our results and
those of \cite{moroi}
indicate no difference.

Analytic expressions may be obtained for the loop functions
if only two mass scales appear.
\ba
J_5(x,1,1,1,1)&=&\frac{1}{6(x-1)^4}\left(1-6x+3x^2+2x^3-6x^2 \ln x
\right)       ~,   \\
& \stackrel {x \ll 1 } \longrightarrow & \frac{1}{6}(1+O(x)) ~, \nonumber \\
J_5(x,x,1,1,1)&=&\frac{1}{2(x-1)^4} \left(1+4 x-5 x^2+2 x(2+x) \ln x
\right)        ~,  \\
& \stackrel {x \ll 1 } \longrightarrow & \frac{1}{2}(1+O(x))  ~, \nonumber \\
J_5(x,x,x,1,1)&=&\frac{1}{2(x-1)^4} \left(-5+4 x+x^2-2 \ln x -4 x \ln x
\right)       ~,   \\
& \stackrel {x \ll 1 } \longrightarrow
& \frac{1}{2}(-5-2 \ln x +O(x)) ~, \nonumber \\
J_5(x,x,x,x,1)&=&\frac{1}{6(x-1)^4} \left(x^2-6 x+3 + \frac{2}{x}
+ 6 \ln x \right)    ~,
\\
& \stackrel {x \ll 1 } \longrightarrow
& \frac{1}{6}(3 + \frac{2}{x} + 6 \ln x +O(x)))
~, \nonumber \\
I_4(x,1,1,1)&=&\frac{1}{2(x-1)^3} \left(-1+x^2 -2 x \ln x \right) ~,
\\
& \stackrel {x \ll 1 } \longrightarrow
 & \frac{1}{2}(1+O(x)) ~, \nonumber \\
I_4(x,x,1,1)&=&\frac{1}{2(x-1)^3} \left(2-2x +(1+x) \ln x \right)  ~,
\\
& \stackrel {x \ll 1 } \longrightarrow
& -\frac{1}{2}(2+ \ln x) ~. \nonumber
\ea
Note that $J_5(x,x,x,1,1)=x^{-2}J_5(x^{-1},x^{-1},1,1,1)$ and
$J_5(x,x,x,x,1)=x^{-2} J_5(x^{-1},1,1,1,1)$, which is
an example of the more general relation in (\ref{scalej}).

The loop functions appearing in the radiative transition dipole
moments involve additional sflavor violating mass squared insertions
beyond those of the flavor conserving dipole moments.
The function
\beq
f^{\prime}_{+,L} =  \frac{m^2_{\tilde{\nu}_L}}{f_+} \frac{\partial f_+}
{\partial m^2_{\tilde{\nu}_L}}
=\frac{\partial \ln f_+}{\partial \ln m^2_{\tilde{\nu}_L}}
~. \nonumber
\eeq
represents the sflavor insertion for the chargino--sneutrino
diagram such
that $f_+  f^{\prime}_{+,L}$ is the loop function.
The loop functions for the sflavor insertions for the three
neutralino diagrams from the first neutralino diagram
of Fig. 2 are described by the same
loop function but with different arguments. These
are represented by a subscript that indicates Bino or Wino
coupling, and right-handed or left-handed sleptons
\ba
f^{\prime}_{\tilde{B}L(R),L (R)} &  = &
\frac{m^2_{\tilde{\ell}_{L(R)}}}{f_{\tilde{B}L(R)}}
\frac{\partial f_{\tilde{B}L(R)}}
{\partial m^2_{\tilde{\ell}_{L(R)}}}=
\frac{\partial \ln f_{\tilde{B}L(R)}}{\partial \ln m^2_{\tilde{\ell}_{L(R)}}}
\nonumber ~, \\
f^{\prime}_{\tilde{W}_0L,L } &  = &
\frac{m^2_{\tilde{\ell}_{L}}}{f_{\tilde{W}_0L}}
\frac{\partial f_{\tilde{W}_0L}}
{\partial m^2_{\tilde{\ell}_{L}}}=
\frac{\partial \ln f_{\tilde{W}_0L}}{\partial \ln m^2_{\tilde{\ell}_{L}}}
\nonumber ~.
\ea
The sflavor violating mass squared
insertions for the second neutralino diagram of Fig. 2
are represented by
\ba
h^{\prime} _{0,L(R)} & = &\frac{m^2_{\tilde{\ell}_{L(R)}}}{h_0}
\frac{\partial h_0}
{\partial m^2_{\tilde{\ell}_{L(R)}}}=
\frac{\partial \ln h_0}{\partial \ln m^2_{\tilde{\ell}_{L(R)}}}
    \nonumber ~, \\
h^{\prime \prime}_{0,LR} & = & { m^2_{\slepton_L} m^2_{\slepton_R} \over h_0}
{\partial^2 ~h_0 \over \partial m^2_{\slepton_L}
                    \partial m^2_{\slepton_R} }
= \frac{1}{h_0} {\partial^2 ~h_0 \over \partial \ln m^2_{\slepton_L}
                    \ln \partial m^2_{\slepton_R} }
\nonumber
\ea
such that $h_0 h^{\prime} _{0,L(R)}$ and $h_0 h^{\prime \prime}_{0,LR}$
are the loop functions.

% ---------------------------------------------------------------


\begin{thebibliography}{99}
\bibitem{cnote} Some of the apparent accidental ``cancellations'' discussed
in the literature are in fact not cancellations among physical
phases, but arise simply from allowed field redefinitions by unphysical phases.
Such misinterpretations of cancellations among phases
may be avoided by working with basis independent combinations
of physical phases.
\bibitem{helicity} Measurement of the final state
lepton helicity, or angular distribution with polarized
initial state, could distinquish the chiral structure of
the transition dipole moment.
\end{thebibliography}
\end{document}